\pdfoutput=1

\documentclass{mfield}

\usepackage{amsmath}

\shorttitle{Sun's Mean Field}
\shortauthors{Sheeley}

\graphicspath{{./}}

\begin{document}

\title{The Sun's Mean Line-of-Sight Field}

\author[0000-0002-6612-3498]{Neil R. Sheeley, Jr.}
\affiliation{Visiting Research Scientist\\
Lunar and Planetary Laboratory, University of Arizona \\
Tucson, AZ 85721, USA}

\begin{abstract}

We regard the Sun-as-a-star magnetic field (\textit{i.e.} the mean field) as a filter for the spherical harmonic components of the photospheric field, and calculate the transmission coefficients of this
filter.  The coefficients for each harmonic, $Y_{l}^{m}$, are listed in three tables according to their
dependence on $B_{0}$, the observer's latitude in the star's polar coordinate system.  These coefficients are used to interpret the 46-yr sequence of daily mean-field measurements at the
Wilcox Solar Observatory.  We find that the non-axisymmetric part of the field originates in the
$Y_{1}^{1}$, $Y_{2}^{2}$, and a combination of the $Y_{3}^{3}$ and $Y_{3}^{1}$ harmonic components.  The axisymmetric part of the field originates in $Y_{2}^{0}$ plus a
$B_{0}$-dependent combination of the $Y_{1}^{0}$ and $Y_{3}^{0}$ components.  The
power spectrum of the field has peaks at frequencies corresponding to the ${\sim}$27-day
synodic equatorial rotation period and its second and third harmonics.  Each of these peaks has fine structure on its low-frequency side, indicating magnetic patterns that rotate slowly under the influence of differential rotation and meridional flow.  The sidebands of the fundamental
mode resolve into peaks corresponding to periods of ${\sim}$28.5 and
${\sim}$30 days, which tend to occur at the start of sunspot maximum, whereas the
$\sim$27-day period tends to occur toward the end of sunspot maximum.  We expect similar rotational sidebands to occur in magnetic observations of other Sun-like stars and to be a useful complement to asteroseismology studies of convection and magnetic fields in those stars.
\end{abstract}


\section{Introduction} \label{sec:intro}

The Sun's mean line-of-sight field is obtained by averaging the 
line-of-sight component of the photospheric magnetic field over the (flat) solar
disk.  The measurement is obtained from Earth, sometimes in integrated sunlight, and is
often called the `Sun-as-a-star' magnetic field, as if the observation were obtained from
the even greater distance of another star. 

 In the early 1970s, John Wilcox proposed to build a new solar telescope in the hills
 south of the Stanford University campus.  The telescope would have relatively
 coarse spatial resolution ${\sim}1$ arcmin and the capability of measuring the Sun's
 mean line-of-sight field.  The idea was not to compete with the telescopes at Mount
Wilson and Kitt Peak, which were already obtaining daily observations of the solar disk
with much higher spatial resolution, but instead to concentrate on factors like sensitivity
and zero-point stability to produce a long-term sequence of relatively precise, global
observations that could be used to study the Sun's large-scale field.
 
Wilcox and Ness had discovered the interplanetary sector structure
using spacecraft data from the Interplanetary Monitoring Platform (IMP)
\citep{1965ICRC....1..302W,1965JGR....70.5793W,1965Sci...148.1592N}.
Also,  by comparing those IMP spacecraft data with photospheric magnetograms
obtained at the Mount Wilson Observatory (MWO), Wilcox and Robert Howard had
shown that the sector structure originated in long-lived, unipolar magnetic
regions on the Sun \citep{1968SoPh....5..564W}.  Consequently, Wilcox thought that
mean field observations would be important for solar-terrestrial studies and, in particular,
would help to improve the ${\sim}$4.5-day timing between the central meridian passage
of a sector boundary at the Sun and at the Earth.  (See the discussion following Douglas
Jones's talk at the Second Solar Wind Conference \citep{1972NASSP.308..122J}.)
As pointed out by \cite{1977SoPh...52D...6S}, Kotov and Severny had already begun daily observations of the mean field at the Crimean Observatory in 1968, and Robert Howard
began them at the Mount Wilson Observatory in 1970.   So Wilcox's interest
in what we now call `space weather' was the motivation for building the Stanford Solar
Observatory\footnote{The observatory was renamed the Wilcox Solar Observatory (WSO)
in 1983 when John Wilcox died while swimming in Mexico}. 

Wilcox's proposal was accepted and daily measurements of the Sun-as-a-star field began
on May 16, 1975.  These measurements were obtained in `integrated sunlight' using a
2.7 m focal length objective lens that creates a 2.5 cm solar image located 3.8 m above
the entrance slit of the spectrograph \citep{1977SoPh...52D...6S}. The new observations quickly confirmed that the strength of the mean
field is correlated with the central-meridian-passage time of low-latitude coronal holes and
with the associated pattern of interplanetary sectors \citep{1976BAAS....8Q.370S,1977SoPh...54..353S,1976SoPh...49..271S}.  In addition, 27-day Bartels displays of the WSO
mean-field measurements matched the corresponding displays of mean field calculated
using the flux-transport model.  This provided one of the first verifications of the transport
model and showed that the mean field originated in flux that spread out from its sources
in active regions  \citep{1985SoPh...98..219S, 1986SoPh..103..203S,1986SoPh..104..425S}.
Many years later, we learned that the mean-field correlates with the occurrence of coronal
inflows seen with white-light coronagraphs on the Solar and Heliospheric Observatory
(SOHO) and Solar Terrestrial Relations Observatory (STEREO) spacecraft
\citep{2015ApJ...809..113S}.

The reason for these correlations is that the mean field is an approximate measure of the
Sun's non-axisymmetric field, and, in particular, of its horizontal dipole and quadrupole
components, $Y_{1}^{1}$ and $Y_{2}^{2}$.  These non-axisymmetric fields are strengthened
by the emergence of flux in active regions, whose bright coronal extensions provide backgrounds
for seeing the much fainter inflows that rain downward from reconnection sites in the outer
corona \citep{ 2017ApJ...835L...7S,2018ApJ...859..135W}.  The outward components of these
reconnections are sometimes observed as streamer blobs moving out through the 30$R_{\odot}$
field of view like `leaves in the wind' and gradually swept up by high speed streams to form
regions of high density \citep{2008ApJ...674L.109S, 2008ApJ...675..853S,2010ApJ...715..300S}.

The purpose of this paper is to analyze an idealized Sun-as-a-star field into its
spherical harmonic components to determine the ones that ought to contribute (both for
the Sun and for a distant star), and then to decompose (or demodulate) the observed WSO
mean field to find out what those contributions have been since the observations began in
1975.   Although unknown to Wilcox in the early 1970s, the extension of these techniques
to other Sun-like stars may complement asteroseismology and exoplanet studies.  

\section{Theoretical Analysis of the Mean Field}\label{sec:harm_components}
\subsubsection{Definition of the  Field}
Let's begin with a definition for the mean line-of-sight field, $B_{m}$.  In general, it is just
the average of the line-of-sight field over the flat solar disk of radius, R:
\begin{equation}
B_{m}~=~\int{B_{los}}dA_{disk}/{\pi}R^2.
\end{equation}
However, we would like to convert to an integral of the radial field, $B_{r}$, over the
surface area $A_{surf}$ of the Sun.  In that case, we need two factors of $\sin{\theta}\cos{\phi}$ --
one factor to convert $B_{los}$ to $B_{r}$, and the other factor to convert $dA_{los}$ to
$dA_{surf}$.  Also, we note that ${\theta}$ and ${\phi}$ are the usual polar and azimuthal
angles in a spherical coordinate system with the $x$-axis pointing toward Earth.  Therefore, Eq(1) becomes
\begin{equation}
B_{m}~=~\int{B_{los}}dA_{disk}/{\pi}R^2~=~\int{B_{r}}(\sin{\theta}\cos{\phi})^{2}dA_{surf}/{\pi}R^2
~=~(1/{\pi})\int_{-{\pi}/2}^{{\pi}/2}\int_{0}^{\pi}{B_{r}}(\sin{\theta}\cos{\phi})^{2}\sin{\theta}d{\theta}d{\phi},
\end{equation}
where ${\theta}$ runs from 0 to ${\pi}$, and ${\phi}$ runs from $-{\pi}/2$ to $+{\pi}/2$.  It is interesting to note that $\sin{\theta}\cos{\phi}$ is the axisymmetric quantity that is usually
called ${\mu}$ in theories of line formation.  So ${\mu}$ is 1 at disk center, 0 at the solar limb,
and $B_{r}$ is heavily weighted toward disk center as
\begin{equation}
B_{m}~=~\int{B_{r}}{\mu}^{2}dA_{surf}/{\pi}R^2.
\end{equation}
As discussed by \cite{1977SoPh...52D...6S}, the weighting toward disk center is even greater
for the WSO observations due to solar limb darkening and diffraction from the entrance
slit of the spectrograph.  In the Appendix of this paper, we consider the limb darkening in detail
and find that the same harmonic components contribute to the mean field for a so-called gray atmosphere in the Eddington approximation as would occur in the absence of limb darkening.  However, the limb darkening reduces the strengths of the $l=1$ and $l=2$ components by 12\%,
and 3.75\%, respectively, and increases the strength of the much weaker $l=3$ components
by 22\%.

Up to this point, I have ignored the $7.25^{\circ}$ tilt of the Sun's axis away from the normal to
the ecliptic plane.  We can include this effect by replacing $\sin{\theta}\cos{\phi}$ with
$\sin{\theta}\cos{\phi}\cos{B_{0}}+\cos{\theta}\sin{B_{0}}$, where $B_{0}$ is Earth's heliolatitude
and varies annually from $-7.25^{\circ}$ in February-March to $+7.25^{\circ}$ in August-September.  In that case, $B_{m}$ becomes
\begin{equation}
B_{m}~=~(1/{\pi})\int_{-{\pi}/2}^{{\pi}/2}\int_{0}^{\pi}{B_{r}}(\sin{\theta}\cos{\phi}\cos{B_{0}}+\cos{\theta}\sin{B_{0}})^{2}\sin{\theta}d{\theta}d{\phi},
\end{equation}
where $B_{r}$ depends on $({\theta},{\phi})$, but $B_{0}$ does not.  Next, our objective
is to expand the binomial factor, $(\sin{\theta}\cos{\phi}\cos{B_{0}}+\cos{\theta}\sin{B_{0}})^2$,
and express $B_{m}$ as the sum of three parts -- one proportional to $\cos^{2}{B_{0}}$,
another proportional to $(\sin{2{B_{0}}})/2$, and the third proportional to $\sin^{2}{B_{0}}$.
\begin{equation}
B_{m}~=~(\cos^{2}{B_{0}})~\frac{1}{{\pi}}\int {B_{r}}(\sin{\theta}\cos{\phi})^{2}d{\Omega}~+~
(\frac{\sin{2{B_{0}}}} {2}) \frac{1}{{\pi}}\int{B_{r}}\sin{2{\theta}}\cos{\phi}~d{\Omega}~+~
(\sin^{2}{B_{0}})~\frac{1}{{\pi}}\int{B_{r}}\cos^{2}{\theta}~d{\Omega},
\end{equation}
where $d{\Omega}=\sin{\theta}d{\theta}d{\phi}$ and the integral sign refers to the double
integral over ${\theta}$ and ${\phi}$, as indicated in Eq(4).

For the Sun, $|B_{0}|~{\leq}~7.25^{\circ}~{\approx}~0.126$ radians so that the
$\sin^{2}{B_{0}}$-factor is ${\leq}0.016$, and can be neglected.  Likewise, the
$\cos^{2}{B_{0}}$-factor is ${\geq}0.984$ and can be replaced by 1.  Finally, the
$(\sin{2{B_{0}}})/2$-factor is approximately $B_{0}$, which will vary between
$-0.126$ and $+0.126$ during the year.  Of course, for a star whose tilt angle is large, we
cannot make these approximations, and we may need to keep all three terms.

\subsubsection{Spherical Harmonic Components}
Next, we consider the form of $B_{r}$.  There are several ways that we could represent this
field.  One way would be to express $B_{r}$ as
a linear combination of the `barber pole' eigenfunctions of the flux-transport equation,
as \cite{1987SoPh..112...17D} did in his theoretical analysis of the Sun's large-scale field.
This approach might help us to interpret the power spectrum of the mean field in terms of
the rigidly rotating patterns that are caused by the latitudinal transport of flux
\citep{1987ApJ...319..481S,1994ApJ...430..399W,1998ASPC..154..131W}.  Another way would
be to represent the field in terms of sectors of the form
$B_{r}=f({\theta})\cos{m}\{{\phi-{\omega}({\theta})t}\}$ (where ${\omega}({\theta})$
is the angular rotation profile), as \cite{1986SoPh..103..203S} did in their analysis of the decay
of the Sun's mean field.  In this paper, I will try a third approach, representing $B_{r}$ as a
linear combination of spherical harmonic components, $Y^{m}_{l}({\theta},{\phi})$, which are
the familiar eigenfunctions of ${\nabla}^{2}B_{r}=0$ on the surface of a sphere.
(See Eqs (9) and (10) below.)
With this understanding, 
\begin{equation}
B_{r}({\theta},{\phi},t)~=~\sum_{l=0}^{\infty}\sum_{m=-l}^{m=l}{{\rho}_{lm}}(t)e^{i{\delta}_{lm}(t)}
Y_{l}^{m}({\theta},{\phi}),
\end{equation}
where $i=\sqrt{-1}$.  Also, ${\rho}_{lm}$ and ${\delta}_{lm}$ are the amplitude and phase of
each harmonic component, $Y_{l}^{m}$, and are defined so that $B_{r}$ is real.   Consequently,
\begin{equation}
B_{m}~=~(\cos^{2}{B_{0}})\sum_{l=0}^{\infty}\sum_{m=-l}^{m=l}{\rho}_{lm}e^{i{\delta}_{lm}}
I_{lm} ~+~(\frac{\sin{2{B_{0}}}} {2})\sum_{l=0}^{\infty}\sum_{m=-l}^{m=l}{\rho}_{lm}e^{i{\delta}_{lm}}J_{lm}~+~
(\sin^{2}{B_{0}})\sum_{l=0}^{\infty}\sum_{m=-l}^{m=l}{\rho}_{lm}e^{i{\delta}_{lm}}K_{lm},
\end{equation}
where the coefficients $I_{lm}$, $J_{lm}$, and $K_{lm}$ are given by

\begin{subequations}
\begin{align}
I_{lm}~=~\frac{1}{{\pi}}\int Y_{l}^{m}({\theta},{\phi}) (\sin{\theta}\cos{\phi})^{2}d{\Omega},\\
J_{lm}~=~\frac{1}{{\pi}}\int Y_{l}^{m}({\theta},{\phi})\sin{2{\theta}}\cos{\phi}~d{\Omega},\\
K_{lm}~=~\frac{1}{{\pi}}\int Y_{l}^{m}({\theta},{\phi})\cos^{2}{\theta}~d{\Omega},
\end{align}
\end{subequations}
and the integral sign refers to the double integral in Eq(4).  Thus, $I_{lm}$, $J_{lm}$, and
$K_{lm}$ (weighted by the respective $B_{0}$-dependent factor)
are real, and indicate the amounts by which the mean-field `filter' reduces the
amplitude ${\rho}_{lm}$ of
the field.  The spherical harmonic functions, $Y_{l}^{m}({\theta},{\phi})$, are defined in terms
of the Associated Legendre functions $P_{l}^{m}(\cos{\theta})$,
 given by \cite{JE_45}, and the normalization factor, $N_{lm}$, as follows
\begin{equation}
Y_{l}^{m}({\theta},{\phi})~=~N_{lm}P_{l}^{m}(\cos{\theta})e^{im{\phi}}
\end{equation}

\begin{equation}
N_{lm}~=~\sqrt {\frac{2l+1}{4{\pi}} \frac{(l-m)!}{(l+m)!}}.
\end{equation}
After some algebra, we can rewrite Eqs(8a)-(8c) as
\begin{subequations}
\begin{align}
I_{lm}~=~\frac{N_{lm}}{{\pi}} \int_{-1}^{1}P_{l}^{m}(x)(1-x^2)dx
\left [ \frac{-4 \sin{(m{\pi}/2)}}{m(m+2)(m-2)} \right ],\\
J_{lm}~=~2\frac{N_{lm}}{{\pi}}  \int_{-1}^{1}P_{l}^{m}(x)x(1-x^2)^{1/2}dx
\left [ \frac{-2 \cos{(m{\pi}/2)}}{(m+1)(m-1)} \right ],\\
K_{lm}~=~\frac{N_{lm}}{{\pi}}  \int_{-1}^{1}P_{l}^{m}(x)x^{2}dx
\left [ \frac{2 \sin{(m{\pi}/2)}}{m} \right ].
\end{align}
\end{subequations}
I used standard Mathematica software \citep{10.5555/320042} to evaluate these
expressions and entered the results in Tables 1-3.  However, Mathematica defines the
Associated Legendre functions using
$P_{l}^{m}(x) = (-1)^{m}(1-x^{2})^{m/2}d^{m}P_{l}(x)/dx^{m}$, which
differ by a factor of $(-1)^{m}$ from the \cite{JE_45} values.  Therefore, I changed
the signs of the odd-m entries in Tables 1-3 to be consistent with the \cite{JE_45} 
convention.  In retrospect, I could have done this automatically by including an extra
factor of $(-1)^{m}$ in each integrand.

Tables 1 and 3 give non-zero contributions to the $Y_{0}^{0}$ magnetic-monopole
component.  Ignoring this term, the main contributions to the $\cos^{2}{B_{0}}$-part
of the mean field come from the $Y_{1}^{1}$, $Y_{2}^{2}$, and $Y_{2}^{0}$ components
with smaller additional contributions from $Y_{3}^{3}$ and $Y_{3}^{1}$.  There are no
contributions when $l=4$ and the higher-order terms are less than 1\%.  Also, for the Sun,
$|B_{0}|~{\leq}~0.126$ and $\cos^{2}B_{0}~{\approx}~1$, so that the contributions of
these harmonics are not weakened appreciably by the $B_{0}$-dependence.

However, in Table 2, the main contributions to the $(\sin{2B_{0}})/2$-part of the mean field
are from the $Y_{1}^{0}$ and $Y_{2}^{1}$ components, which are antisymmetric across
the equator.   These contributions are +0.244 and
+0.206, respectively.
For small $B_{0}$, $(\sin{2}{B_{0}})/2~{\approx}~B_{0}$,
which varies annually from -0.126 to +0.126.  Consequently, the expected mean-field
contributions of the $Y_{1}^{0}$ and $Y_{2}^{1}$ components vary annually and have
peak amplitudes
of $\pm$3.1\% and $\pm$2.6\%, respectively.  These contributions are comparable
to the relatively small, but finite, 3.5\% and 2.7\% contributions of the $Y_{3}^{3}$ and
$Y_{3}^{1}$ components in Table 1.  Thus, they ought to be  noticeable, especially
during sunspot cycles when the polar fields are strong.  The other contributions from
Table 2 are comparable to the contributions of the higher-order terms in Table 1, which
we have already chosen to neglect.

In Table 3,  $K_{20}=+0.168$ and $K_{31}=+0.121$.  However, these relatively large
values can be ignored because the $\sin^{2}{B_{0}}$ factor ($~{\sim}~0.016$)
reduces their net mean-field contributions to less than 1\%.  Although these entries
in Table 3 are unimportant for the Sun, they might contribute appreciably
to the mean-field of other stars whose rotation axes may be directed closer to the line
of sight.  (Of course, for those distant stars, the annual variations induced
by Earth's motion around the Sun would be negligible.) 
\begin{table}[h!]
\caption{Elements of $I_{lm}$ \{for the $\cos^{2}B_{0}$-term\}}
\begin{center}
\begin{tabular}{c c c c c c c c c}
\hline\hline
$l/m$ & 0&1 & 2 & 3 & 4 & 5 & 6 & 7 \\[0.5ex]
\hline\
0 &+0.188 & \\[1.5ex]

1 & 0 & +0.173  \\[1.5ex]

2 & -0.084 & 0 & +0.103	\\[1.5ex]

3 & 0 & -0.027 & 0 & +0.035	 \\[1.5ex]

4 & 0 & 0 & 0 & 0 & ~~0~~  \\[1.5ex]

5 & 0 & -0.003 & 0 & +0.004 &0 & -0.005  \\[1.5ex]

6 & 0 & 0 & 0 & 0 & 0 & 0 & ~~0~~  \\[1.5ex]

7 & 0 & -0.001 & 0 & +0.001 & 0 & -0.001 & 0 & +0.002  \\[1.5ex]
\hline 
\end{tabular}
\end{center}
\end{table} 
\begin{table}[h!]
\caption{Elements of $J_{lm}$ \{for the $(\sin{2B_{0}})/2$-term\}}
\begin{center}
\begin{tabular}{c c c c c c c c c}
\hline\hline
$l/m$ & 0&1 & 2 & 3 & 4 & 5 & 6 & 7 \\[0.5ex]
\hline\
0 & 0 & \\[1.5ex]

1 & +0.244 & 0  \\[1.5ex]

2 & 0 & +0.206 & 0	\\[1.5ex]

3 & -0.093 & 0 & +0.085 & ~~0~~	 \\[1.5ex]

4 & 0 & 0 & 0 & 0 & 0  \\[1.5ex]

5 & -0.018 & 0 & +0.018 & 0 & -0.015 & ~~0~~  \\[1.5ex]

6 & 0 & 0 & 0 & 0 & 0 & 0 & 0  \\[1.5ex]

7 & -0.007 & 0 & +0.007 & 0 & -0.007 & 0 & +0.006 & ~~0~~  \\[1.5ex]
\hline 
\end{tabular}
\end{center}
\end{table} 

\begin{table}[h!]
\caption{Elements of $K_{lm}$ \{for the $\sin^{2}B_{0}$-term\}}
\begin{center}
\begin{tabular}{c c c c c c c c c}
\hline\hline
$l/m$ & 0&1 & 2 & 3 & 4 & 5 & 6 & 7 \\[0.5ex]
0 & +0.188 & \\[1.5ex]

1 & 0 & +0.086  \\[1.5ex]

2 & +0.168 & 0 & ~~0~~	\\[1.5ex]

3 & 0 & +0.121 & 0 & -0.017  \\[1.5ex]

4 & 0 & 0 & 0 & 0 & ~~0~~  \\[1.5ex]

5 & 0 & +0.045 & 0 & -0.034 & 0 & +0.007  \\[1.5ex]

6 & 0 & 0 & 0 & 0 & 0 & 0 & ~~0~~  \\[1.5ex]

7 & 0 & +0.026 & 0 & -0.023 & 0 & +0.017 & 0 & -0.004  \\[1.5ex]
\hline
\end{tabular}
\end{center}
\end{table} 

In summary, the Sun-as-a-star field is dominated by the $Y_{1}^{1}$, $Y_{2}^{2}$, and
$Y_{2}^{0}$ components of the Sun's field.  Also, the mean field has much smaller contributions
from the $Y_{3}^{3}$ and $Y_{3}^{1}$ components, and from the $Y_{1}^{0}$ and $Y_{2}^{1}$ components whose strengths are modulated by the annual variation of $B_{0}$ as Earth
orbits the Sun.  Finally, it is important to realize
that we have been describing the `transmission factors' of the mean-field filter and that
the real mean field also depends on the amplitude, ${\rho}_{lm}$, and phase,
${\delta}_{lm}$, of the radial field, $B_{r}$, that is being filtered.   Next, we will use these results
to interpret mean-field observations obtained daily at the Wilcox Solar Observatory
during the 46-yr interval from May 16, 1975 to the present.
\clearpage
\section{Sun-as-a-Star Magnetic Field Measurements from WSO}
\subsection{27-day rotational modulation}
Figure~1 shows daily measurements of the Sun's mean field obtained at WSO since
16 May 1975.  Approximately
\begin{figure}[h!]
 \centerline{
 \fbox{\includegraphics[bb=110 420 500 673,clip,width=0.65\textwidth]
 {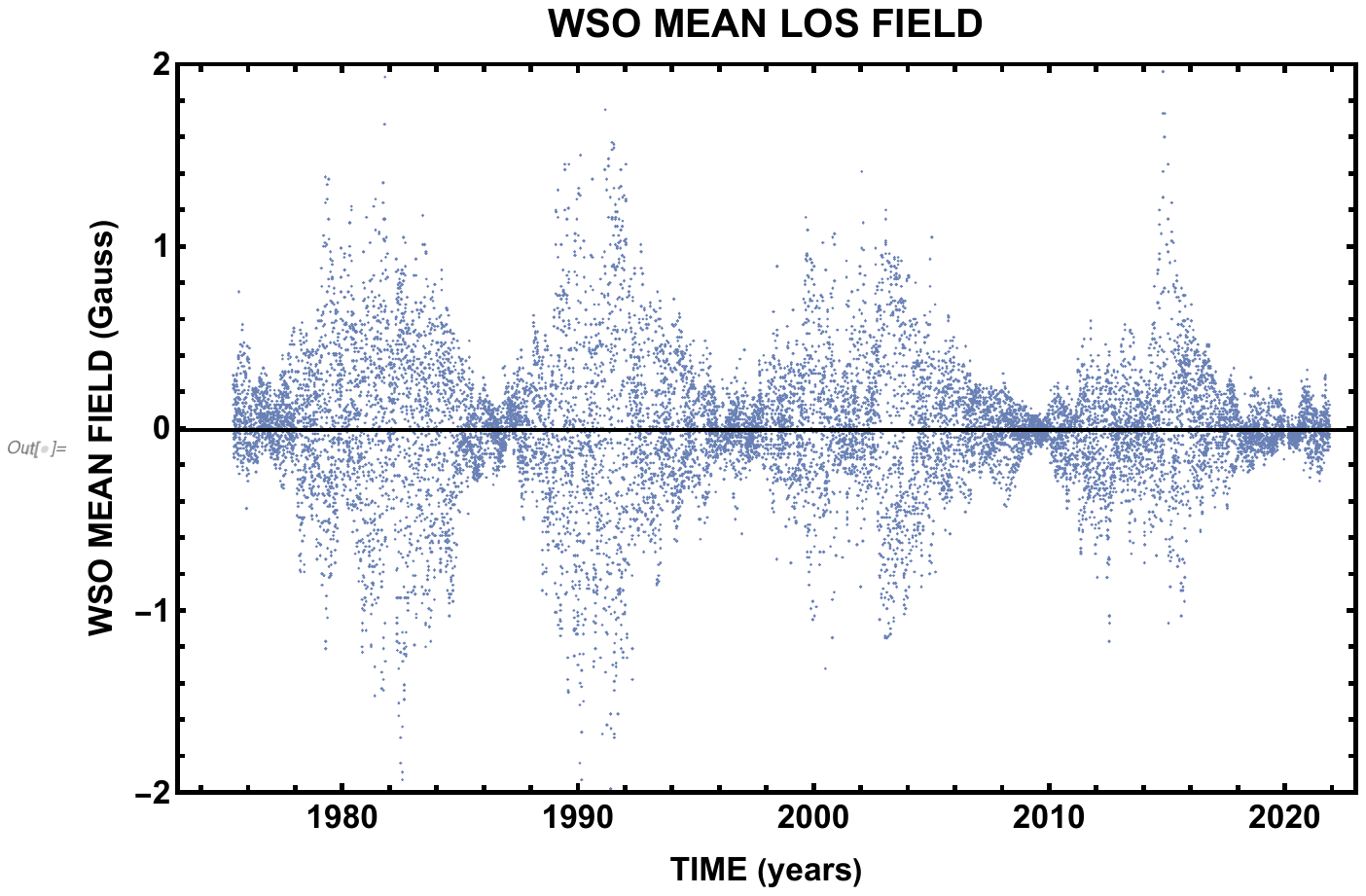}}}
\caption{Daily measurements of the Sun's mean line-of-sight field obtained at WSO from 16 May 1975 to 16 November 2021.  The blurred cloud of points shows
peaks of strength ${\gtrsim}$1G in each cycle, and valleys of
strength ${\lesssim}$0.2G near sunspot minimum.
\label{fig:fig1}}
\end{figure}
\noindent
17,000 points over 46 years give a blurred distribution
with peaks and valleys around sunspot maximum and minimum
in each of four sunspot
cycles.  This figure is essentially the same as the one that is shown on the WSO web
site (http://wso.stanford.edu), and leaves us with the question of how to extract information
from these data.
A big clue is contained in a plot of the data obtained during the first year of observations,
as shown in Figure~2.  The field
\begin{figure}[h!]
 \centerline{
 \fbox{\includegraphics[bb=110 440 480 673,clip,width=0.65\textwidth]
 {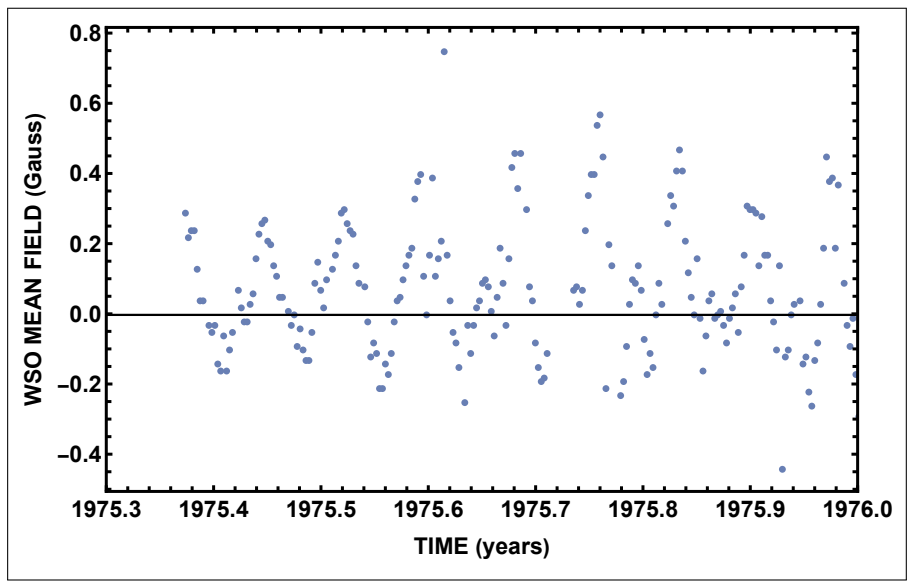}}}
\caption{Mean-field measurements during the first year of observations at WSO,
showing the end of a 27-day recurrence pattern, that was associated with
the gradual demise of a long-lived coronal hole.
\label{fig:fig2}}
\end{figure}
\noindent
oscillates with a period of about 27 days (${\sim}$0.074 yr),
before degrading toward the end of the year, as the corresponding low-latitude coronal
hole gradually died \citep{1976SoPh...49..271S}.
What we
would like to know is how much power is contained in this 27-day modulation and
how that power varies with time during the 46-yr interval.  In effect, we are asking for
the envelope of this mean-field time series. 

There are several approximately equivalent ways to produce this envelope.  One way is
to divide the time base into 27-day segments and to compute the maximum-minimum
difference of the mean-field values on each segment.  By plotting the absolute values of 
these differences, we would obtain a display of the mean field variation similar to the one
that \cite{2015ApJ...809..113S} obtained in their paper describing the rejuvenation of
the Sun's large-scale magnetic field.  A similar result is obtained by computing the
standard deviation of the mean field on each 27-day segment (allowing for data
gaps when taking the averages), and then plotting that value as a function of time.
Another, nearly equivalent procedure is to set the mean-field data gaps equal to zero before computing the standard deviations, and then to perform a 27-day moving average
of those standard deviations.  This is the approach that I shall use in the remainder of this paper.

Figure~3 shows this 27-day moving average, compared
with the monthly averaged sunspot number (divided by 2000) during cycles 21 - 24.  As
described previously using the `max-min' display \citep{2015ApJ...809..113S}, 
the mean field originates in episodic bursts whose amplitudes often tend to be large as the
sunspot cycle enters its declining phase.  Also, one can see the decrease of the mean
field strength during 1976 as the sunspot number reached its 11-year minimum and
the low-latitude coronal hole died (\textit{cf.} Figure~2).  Finally, note that the amplitudes
of the peaks in Figure~3 are about 1.4 times smaller than those in Figure~1.  This is close to the
${\sqrt{2}}$ that one might expect
for the difference between the standard deviation and the envelope of a curve.  (For example,
if $f(t)=A(t)cos{\omega}t$, the envelope is ${\pm}A(t)$ and the standard deviation is
$A(t)/\sqrt{2}$.)
\begin{figure}[h!]
 \centerline{
 \fbox{\includegraphics[bb=110 425 500 673,clip,width=0.64\textwidth]
 {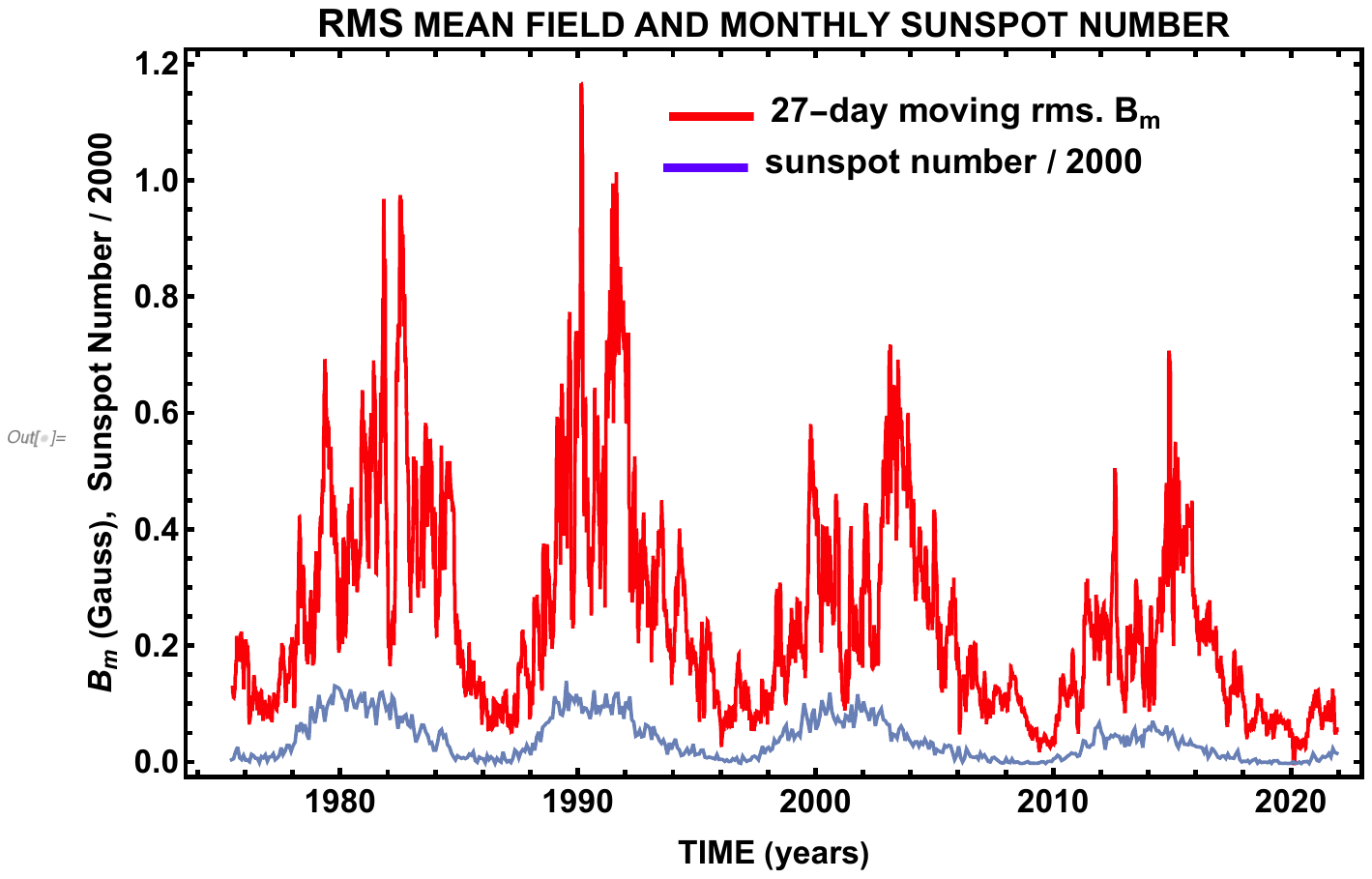}}}
\caption{Root-mean-square deviation of the WSO mean field from its 27-day average value (red), compared with the monthly averaged sunspot number from the Royal Observatory of Belgium (SILSO) divided by 2000 (blue) during cycles 21 - 24.  This comparison shows the tendency of
the mean field to occur in episodic bursts, often when the sunspot cycle begins its decline.
\label{fig:fig3}}
\end{figure}

\subsection{Using the Fourier transform approach}
Next, we return to the WSO mean-field measurements that we displayed as a function of time
in Figure~1.  After setting the missing field strengths equal to 0, we take the discrete Fourier transform defined by
\begin{equation}
f_{s}~=~\frac{1}{{\sqrt{N}}}\sum_{k=1}^{N}{B_{k}}e^{2{\pi}i(k-1)(s-1)/N}
\end{equation}
 where $B_{k}$ refers to the individual mean-field measurements whose index, $k$, runs from
 1 to $N=16,987$, corresponding to the most recent measurement on 16 November 2021.
 In this
 case, the frequency, ${\omega}$, in rad $\textrm{day}^{-1}$ is given by
 \begin{equation}
{\omega}=2{\pi}s/N. 
 \end{equation}
 Because $f_{s}$ is a complex
 number, one typically plots the power, $P({\omega})$, defined as the positive number,
$f_{s}^{*}f_{s}$, versus ${\omega}$.  However, to keep the units in Gauss, I plot
 the positive square root of this power.  Also, it is not necessary to include the full
 range $(0,2{\pi})$ because the spectrum is symmetric about the point ${\omega}={\pi}$.
Moreover, we do not even need to include all of the half range $(0,{\pi})$ because the lines
disappear after the $m=3$ peak around ${\omega}=0.7$ rad $\textrm{day}^{-1}$.
This is consistent with our expectations from Table 1, which gives 0 for $m=4$,
and less than 1\% for  $m=5$.  

The power spectrum in Figure~4 shows three main peaks at frequencies
of approximately ${\omega}=0.231$, 0.460, and 0.702 $\textrm{rad~day}^{-1}$.
These frequencies are in the ratio of approximately 1:2:3, corresponding to a
fundamental rotation rate
\begin{figure}[h!]
 \centerline{
 \fbox{\includegraphics[bb=110 412 515 668,clip,width=0.69\textwidth]
 {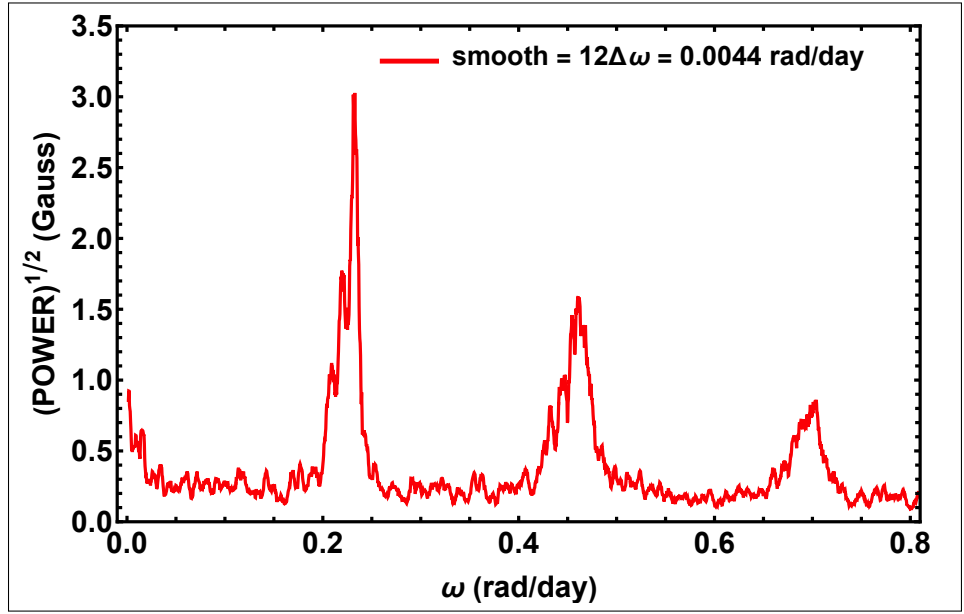}}}
\caption{Power spectrum of daily WSO mean-field measurements during 1975 - 2021,
showing three main sectorial peaks
at ${\omega}~{\approx}~0.231$, 0.460, and 0.702 rad $\textrm{day}^{-1}$
(corresponding to a rotation period of 27.2 days and its second and third harmonics).   The
plot has been smoothed in a moving average of 12 resolution elements, each of size
${\Delta}{\omega}=2{\pi}/N=3.70{\times}10^{-4}$ rad $\textrm{day}^{-1}$.
\label{fig:fig4}}
\end{figure}
\noindent
with $m=1$ and its first two harmonics with $m=2$ and
$m=3$.  The associated periods are approximately 27.2, 13.6, and 9.0 days, respectively.
Evidently, we are seeing rigidly rotating recurrence patterns of two-sector, four-sector, and
six-sector fields (\textit{i.e.} the dipole, quadrupole, and hexapole fields).

These regions are shown separately in the three panels of Figure~5.  Each spectrum is displayed with the same 12 unit
\begin{figure}[h!]
 \centerline{
 \fbox{\includegraphics[bb=115 422 490 665,clip,width=0.31\textwidth]
 {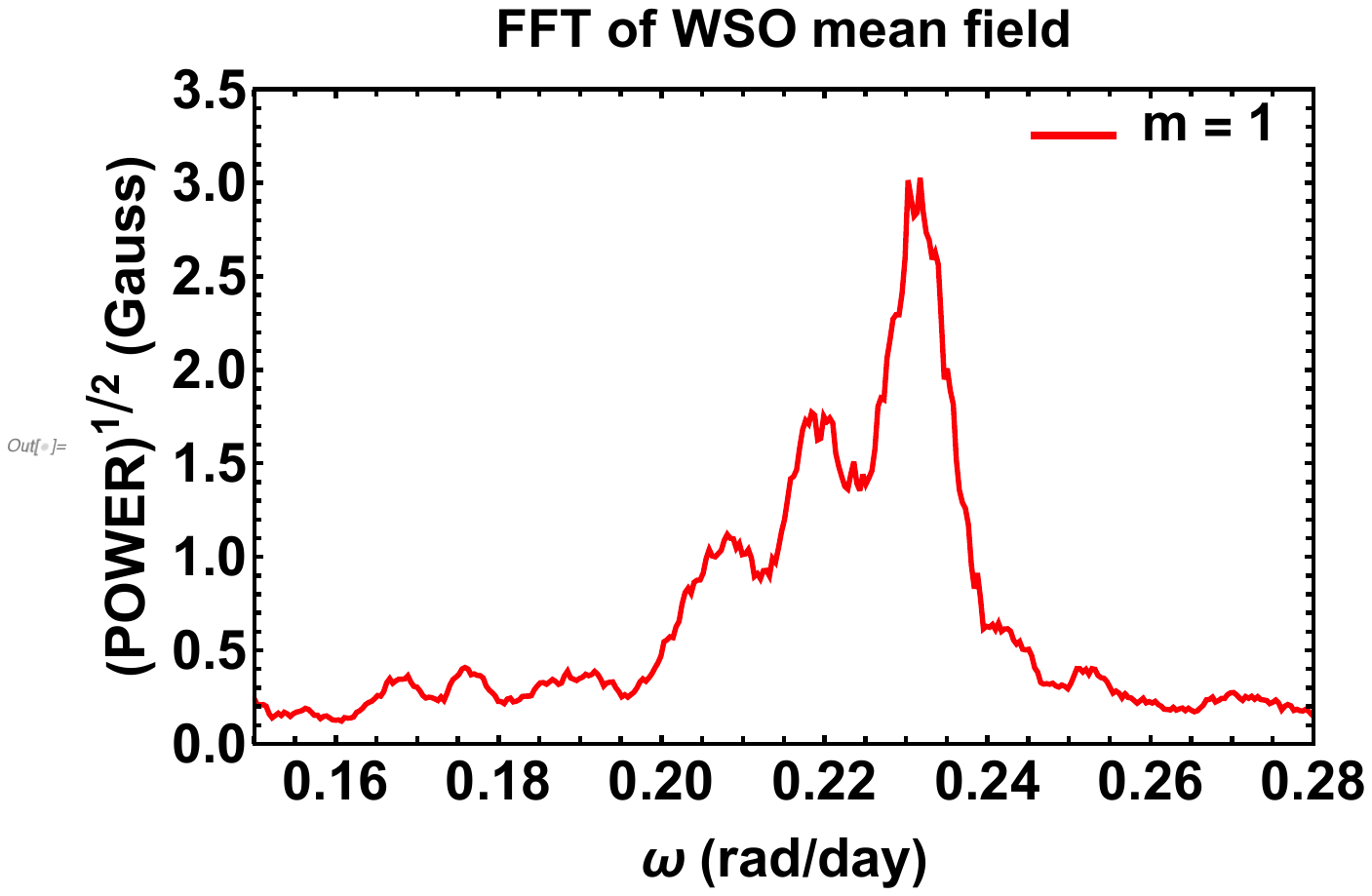}}
 \hspace{0.01in}
  \fbox{\includegraphics[bb=115 413 495 665,clip,width=0.30\textwidth]
 {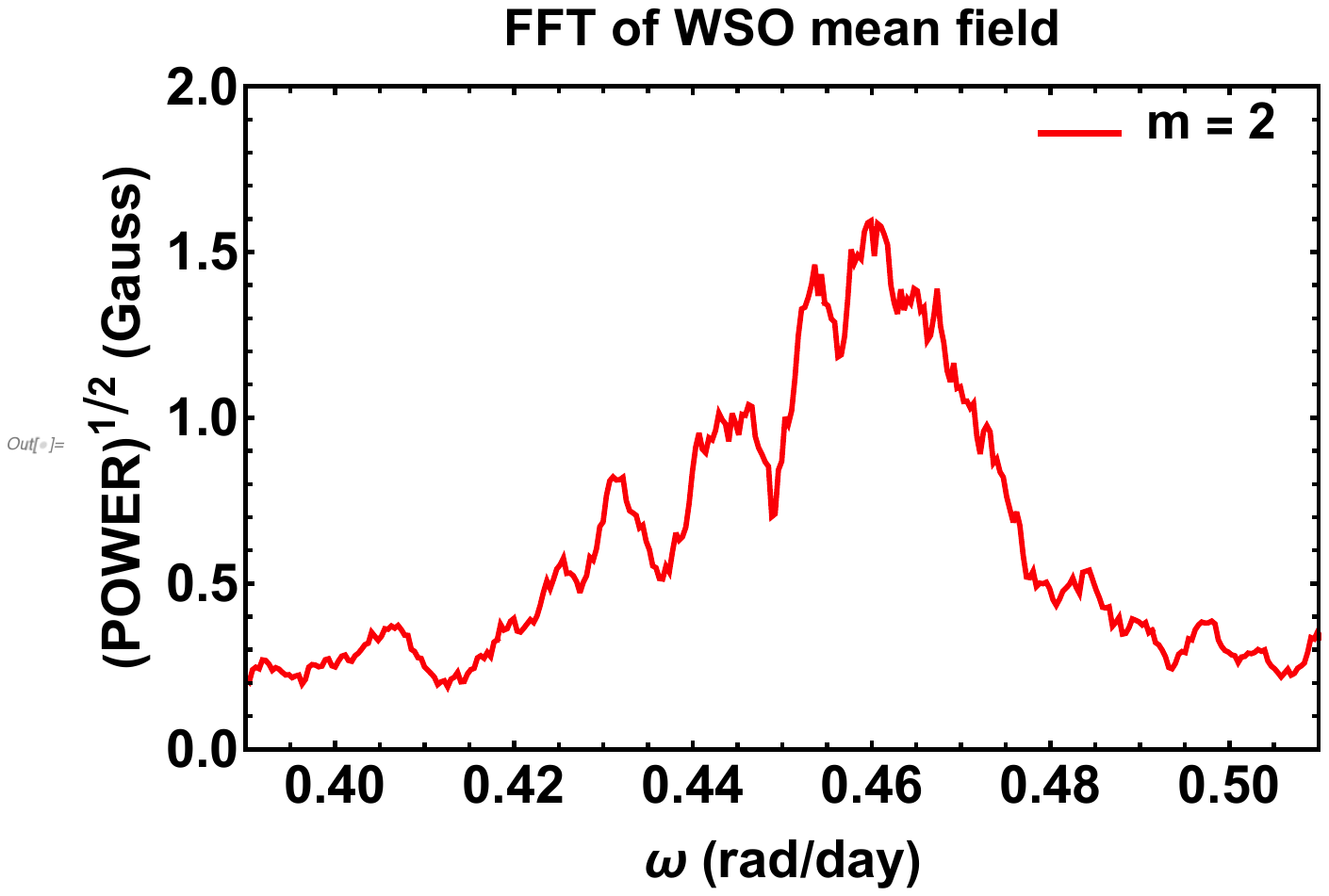}}
  \hspace{0.01in}
  \fbox{\includegraphics[bb=115 422 490 665,clip,width=0.31\textwidth]
 {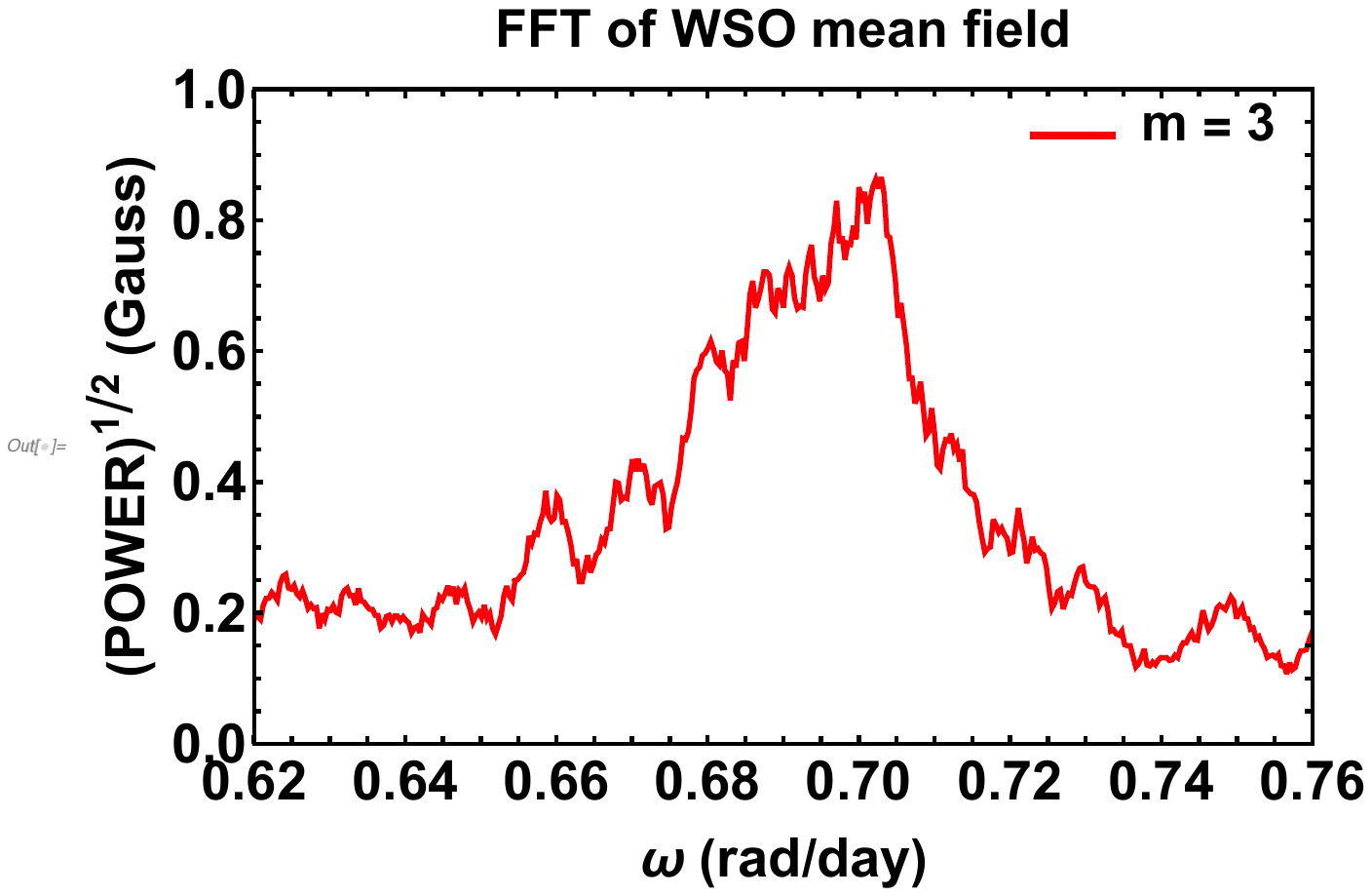}}} 
\caption{Segments of the power spectrum in Figure~4, showing the
fine structure of the $m=1$ (left), $m=2$ (middle), and $m=3$ (right)
components for the full time interval 1975 - 2021.
\label{fig:fig5}}
\end{figure}
\noindent
smoothing, which is
12 times the $2{\pi}/N$ resolution for $N{\sim}$ 16,987 points, and corresponds to
0.0044 $\textrm{rad~day}^{-1}$.  On the
low-frequency side of
the fundamental peak at ${\omega}=0.231~ \textrm{rad~day}^{-1}$, there are
peaks at 0.219 and 0.208 $\textrm{rad~day}^{-1}$, corresponding to periods
of approximately 28.7 and 30.2 days, respectively.  The 28.7-day period is comparable
to the ${\sim}$28.5-day recurrence period of the slanted patterns seen around sunspot
maximum in Bartels displays of the interplanetary magnetic field observed by in-ecliptic
spacecraft and inferred from Earth-based magnetometers 
\citep{1975SoPh...41..461S, 1984PhDT.........5H,1985SoPh...98..219S,1986SoPh..104..425S,1994JGR....99.6597W}.  However, the 30.2-day peak has no in-ecliptic counterpart, and
therefore probably originates in rigidly rotating structures at latitudes that are beyond the
reach of the in-ecliptic measurements, as discussed in Section 3.2.1 below.

Although the frequencies of the main peaks of the $m=1$, $m=2$, and $m=3$ distributions
occur in the ratio of 1:2:3, the frequencies of the sidebands do not occur in this ratio.  In particular, the $m=2$ and $m=3$ sidebands are not `blurred out' harmonics of the peaks at 0.208 and
0.219 $\textrm{rad~day}^{-1}$.  Not only do the $m=2$ structures occur at different
frequencies than we would expect for second harmonics, but also these structures
are accompanied by additional features for which there is
no corresponding peak in the sidebands of $m=1$.   For $m=3$, there are even more
fluctuations, crowding into a broad slope of nearly continuous intensity.

Finally, we note in Figure~4 that there is a noisy `ledge' of strength
${\sim}$0.6 G at ${\omega}~{\sim}~0.017~\textrm{rad~day}^{-1}$.  This structure
corresponds to a weak annual variation associated with the motion of the Earth around
the Sun.  As discussed in Section 2, this annual variation is introduced through
$B_{0}$ - Earth's latitude in the Sun's polar coordinate system.  At even lower
frequencies, the spectrum rises steeply, and the expected peak
at ${\omega}~{\approx}~0.0016~\textrm{rad~day}^{-1}$ (corresponding to the
11-yr sunspot cycle) is not visible in this $0.0044~\textrm{rad~day}^{-1}$ smoothed
plot.  \cite{2019Ap&SS.364...45K} has used mean-field observations since 1968 to
study this annular variation in greater detail.

\subsubsection{The temporal origin of the peaks in the power spectrum}
The next problem is to find the temporal origin of these spectral peaks.
We do this by selecting the frequency range of interest and then taking the inverse
Fourier transform through that spectral window.  We use the inverse transform
\begin{equation}
B_{k}~=~\frac{1}{{\sqrt{N}}}\sum_{s=1}^{N}{f_{s}}e^{-2{\pi}i(k-1)(s-1)/N},
\end{equation}
but with $f_{s}$ multiplied by a function of $s$ (or equivalently ${\omega}$) that is 1
on the interval of interest and
0 elsewhere.  Note that $f_{s}$ is the original complex Fourier transform given by Eq(12),
and not the absolute value that was used in Figure~4.  In general, the inverse Fourier
transform is also a complex number, so we calculate the standard deviation, ${\sigma}$, from
the relation ${\sigma}^2 =<|B_{k}|^2>-~|<B_{k}>|^2$, where in the second term, we compute
the 27-day average of $B_{k}$ before we take its absolute value and square it.  In this case,
it is easy to show that ${\sigma}^2={\sigma}_{r}^2+{\sigma}_{i}^2$ (\textit{i.e.} the standard deviation of $B_{k}$ is the square root of the sum of the squares of the standard deviations
of its real and imaginary parts).

Figure~6 was created by selecting a disjoint interval consisting of the three principal peaks of the
power spectrum
 \begin{figure}[h!]
 \centerline{
 \fbox{\includegraphics[bb=110 405 517 670,clip,width=0.64\textwidth]
 {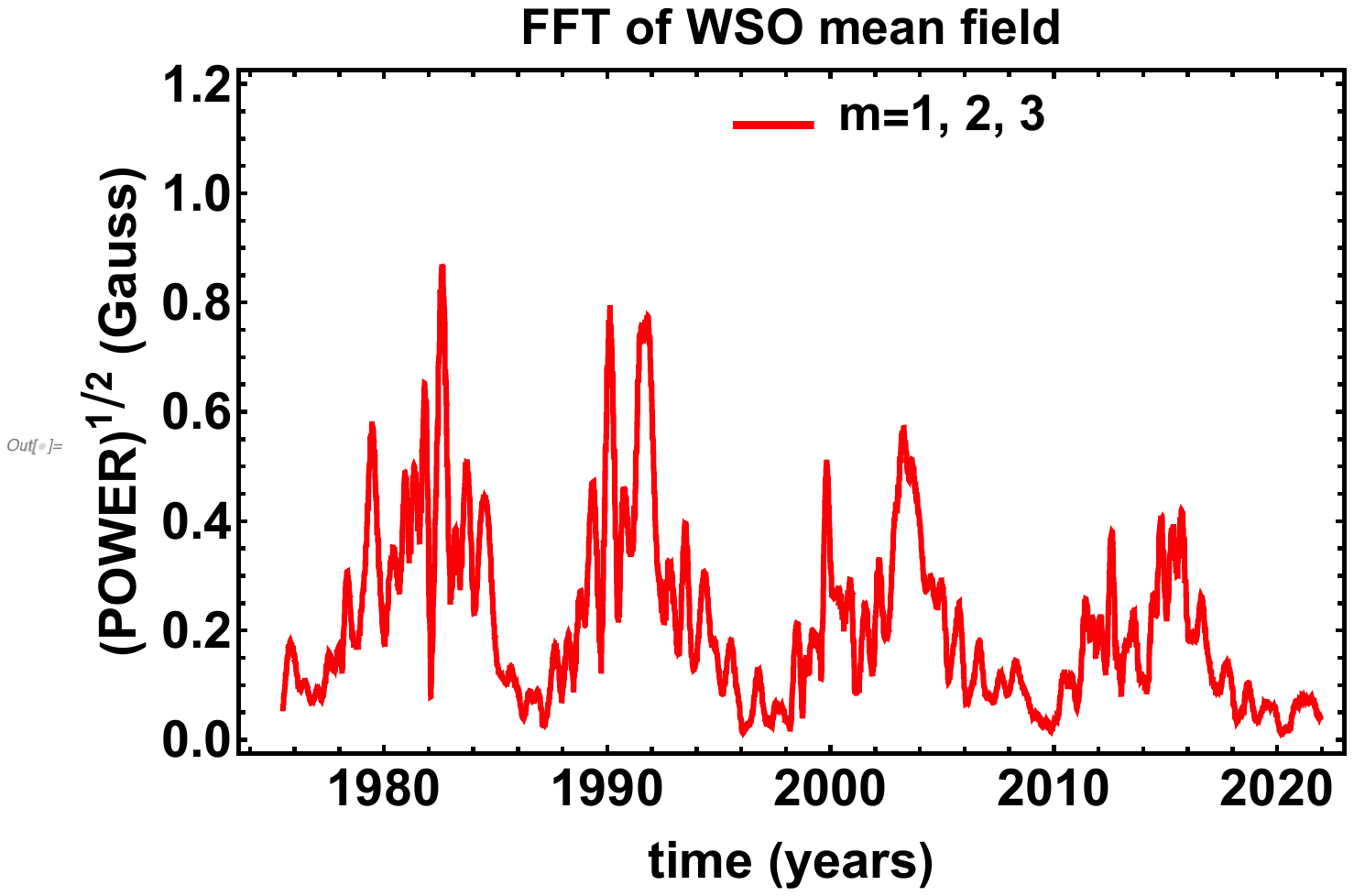}}}
\caption{Result of inverting the Fourier transform of the mean field back to the
temporal domain using all three frequency windows (0.20-0.25), (0.42-0.50), and (0.65-0.73)
$\textrm{rad~day}^{-1}$ (corresponding to the $m=1$, $m=2$, and $m=3$ sectoral modes,
respectively), and  then displaying the 27-day moving standard deviation of this inverted transform.
\label{fig:fig6}}
\end{figure}
\noindent
 - specifically,
 (0.20-0.25), (0.42-0.50), and (0.65-0.73) $\textrm{rad~day}^{-1}$ - and then by displaying
the root-mean-square power in the inverse
transform.  This plot is essentially the same as Figure~3 without
the noise.  Because
we have not included power at the frequency of the annual
variation (0.0172 $\textrm{rad day}^{-1}$), we have removed potential contributions
from Table 2, so that the only contributions come from modes in Table 1.  This leaves only
the $Y_{1}^{1}$ (and possibly the $Y_{3}^{1}$) components as likely contributors from the
$m=1$ sector, and only the $Y_{2}^{2}$ and $Y_{3}^{3}$ components as contributors from
the $m=2$ and $m=3$ sectors, respectively.   So, the red curve in Figure~6 indicates
contributions from the horizontal dipole, quadrupole, and hexapole components
($Y_{1}^{1}$, $Y_{2}^{2}$, and $Y_{3}^{3}$), and possibly a small contribution from
$Y_{3}^{1}$. 

Next, we ask how this non-axisymmetric power is distributed among the $m=1$, $m=2$,
and $m=3$ sectorial modes.  For this purpose, we use the individual frequency ranges
(0.20-0.25), (0.42-0.50), and (0.65-0.73) $\textrm{rad~day}^{-1}$, which
correspond to the fundamental, second, and third harmonics shown in  Figures~4 and 5.
The 27-day running 
\begin{figure}[b!]
 \centerline{
 \fbox{\includegraphics[bb=110 405 517 670,clip,width=0.60\textwidth]
 {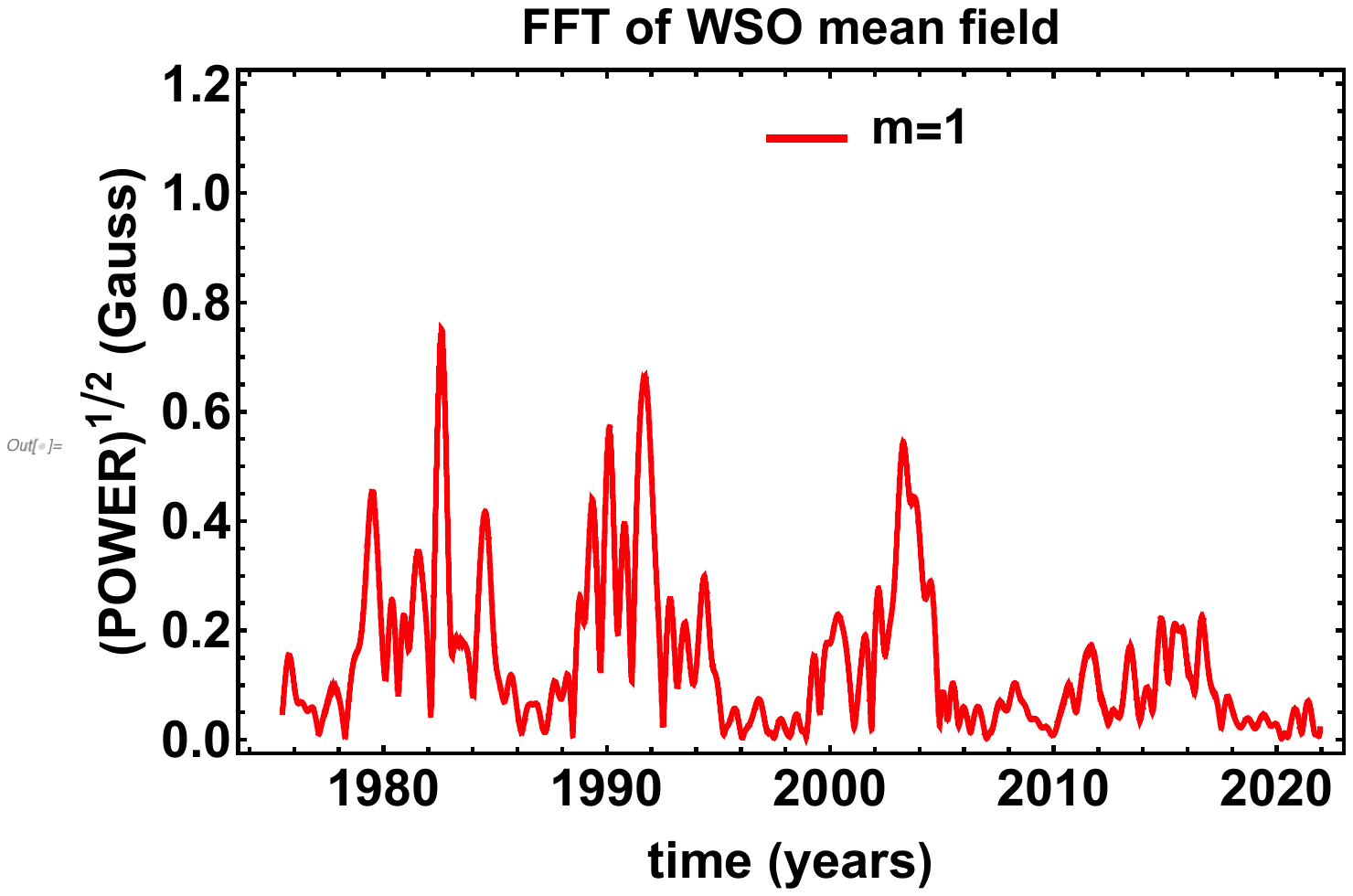}}}
\centerline{
 \vspace{0.01in}
  \fbox{\includegraphics[bb=110 405 517 670,clip,width=0.60\textwidth]
 {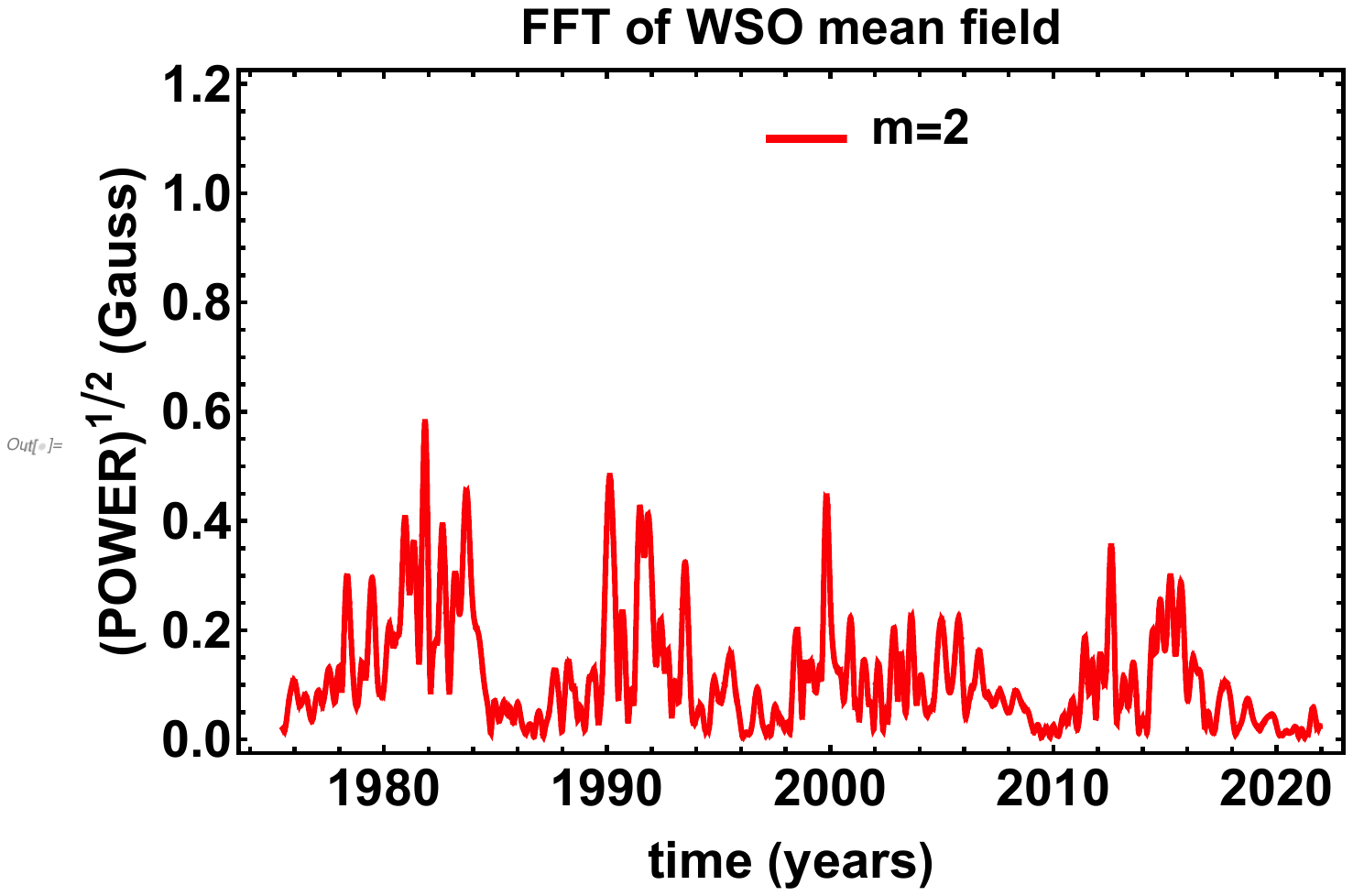}}}
 \centerline{
 \vspace{0.01in}
  \fbox{\includegraphics[bb=110 405 517 670,clip,width=0.60\textwidth]
 {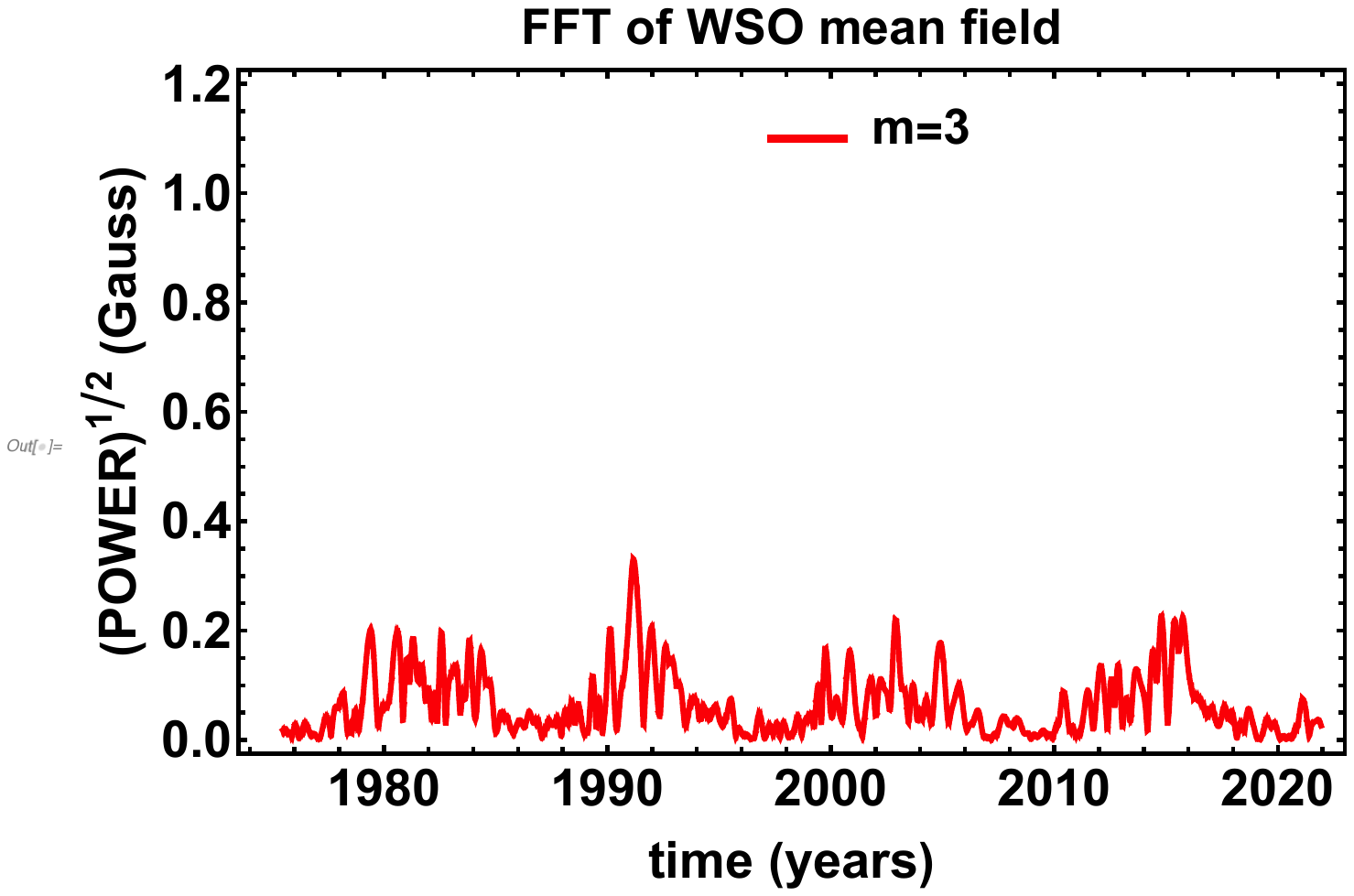}}}
\caption{Power in the two-sector (top), four-sector (middle), and six-sector (bottom)
patterns of the mean field.  Referring to Table~1, we expect this power to originate
mainly in the $Y^{1}_{1}$, $Y^{2}_{2}$, and $Y^{3}_{3}$ components of
the field.
\label{fig:fig7}}
\end{figure}
\noindent
averages are shown in Figure~7.  In general, the $m=1$ component
contributes the most and the $m=3$ component
contributes the least.   With a few exceptions, the mean-field is
dominated by the $m=1$ and $m=2$ sectoral modes.  The $m=1$ component has
large peaks in 1982, 1991, 2003, and a small one in 2015 that we recall
from a nearly identical plot of the equatorial dipole, that was derived from spatially
resolved WSO observations \citep{ 2015ApJ...809..113S}.
Also, the $m=2$ component has moderately large, but narrow, peaks in 1981, 2000,
and 2012.  The $m=3$ component has a large, narrow peak in 1991 when the
$m=1$-2 values are temporarily low.  Also, in the relatively weak sunspot cycle 24,
the three components have nearly coincident peaks of approximately equal strength,
which combine to give a stronger peak of total mean-field power, as seen in Figure~6.

Based on the factors in Tables 1 and 2, we expect that the
$m=1$ power
originates primarily from the $Y_{1}^{1}$ horizontal dipole component of field and 
secondarily from the $Y_{3}^{1}$ component.  As mentioned above,
by filtering out annual variations, we have excluded contributions from the $Y_{2}^{1}$
component, that would otherwise occur through the $B_{0}$ factor.  Likewise, the 
$Y_{3}^{2}$ component is excluded from the $m=2$ sector, which indicates power in the
$Y_{2}^{2}$ component alone.

Finally, we have the impression in Figure~7 that the temporal fluctuations become
systematically finer as $m$ increases from 1 to 3.  This is consistent with the increased
coarseness of the power spectra in Figures~4 and 5 as $m$ increases:  For $m=1$, the peaks
were often resolvable; for $m=2$, they formed coarser structures; and for $m=3$,
they merged into a `bumpy' continuum.  In a previous analytical study of the mean field,
we noted that the field depended on the product $mt$, not on $m$ and $t$ separately.
This meant that the mean field would decay as $1/m$ in the absence of meridional flow
 \citep{1986SoPh..103..203S}.  In particular, a 4-sector field would decay twice as fast as
 a 2-sector field, and a 6-sector field would decay three times as fast.  When flow was
 present, this analytical simplification did not occur.  However, the apparent trend in Figure~7
 indicates that a monotonic relation may still be present.

Next, we look for the origins of the three $m=1$ peaks at
${\omega}=0.231$, 0.219, and~0.208 $\textrm{rad day}^{-1}$, shown in the left panel
of Figure~5.  To do this, we select relatively narrow spectral windows
surrounding these peaks, and then invert the Fourier transform and plot the running
27-day rms averages.  For the intervals, ${\omega} = 0.226-0.237$,
${\omega} = 0.215-0.225$
\begin{figure}[h!]
 \centerline{
 \fbox{\includegraphics[bb=110 405 517 670,clip,width=0.60\textwidth]
 {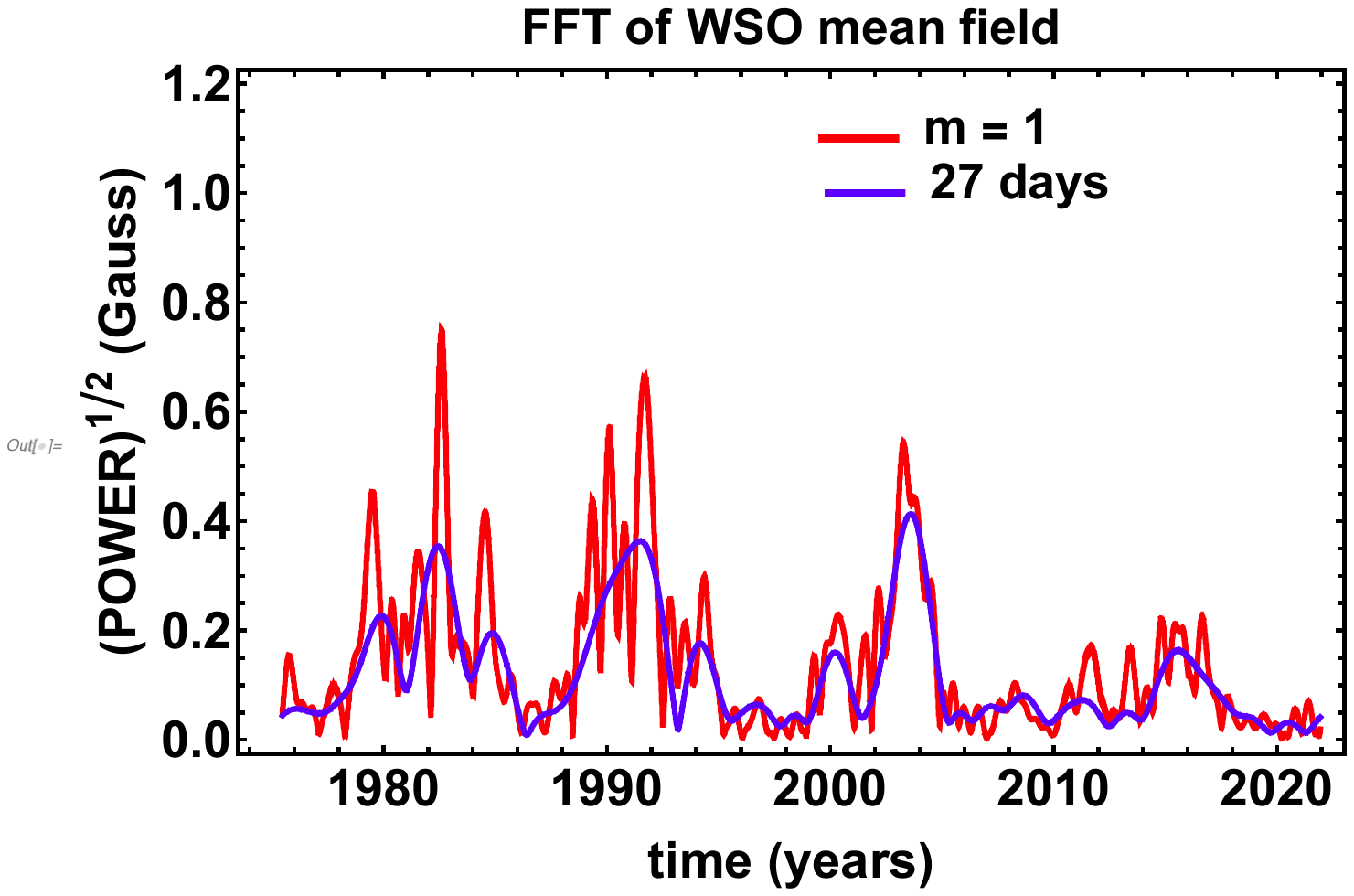}}}
\centerline{
 \vspace{0.01in}
  \fbox{\includegraphics[bb=110 405 517 670,clip,width=0.60\textwidth]
 {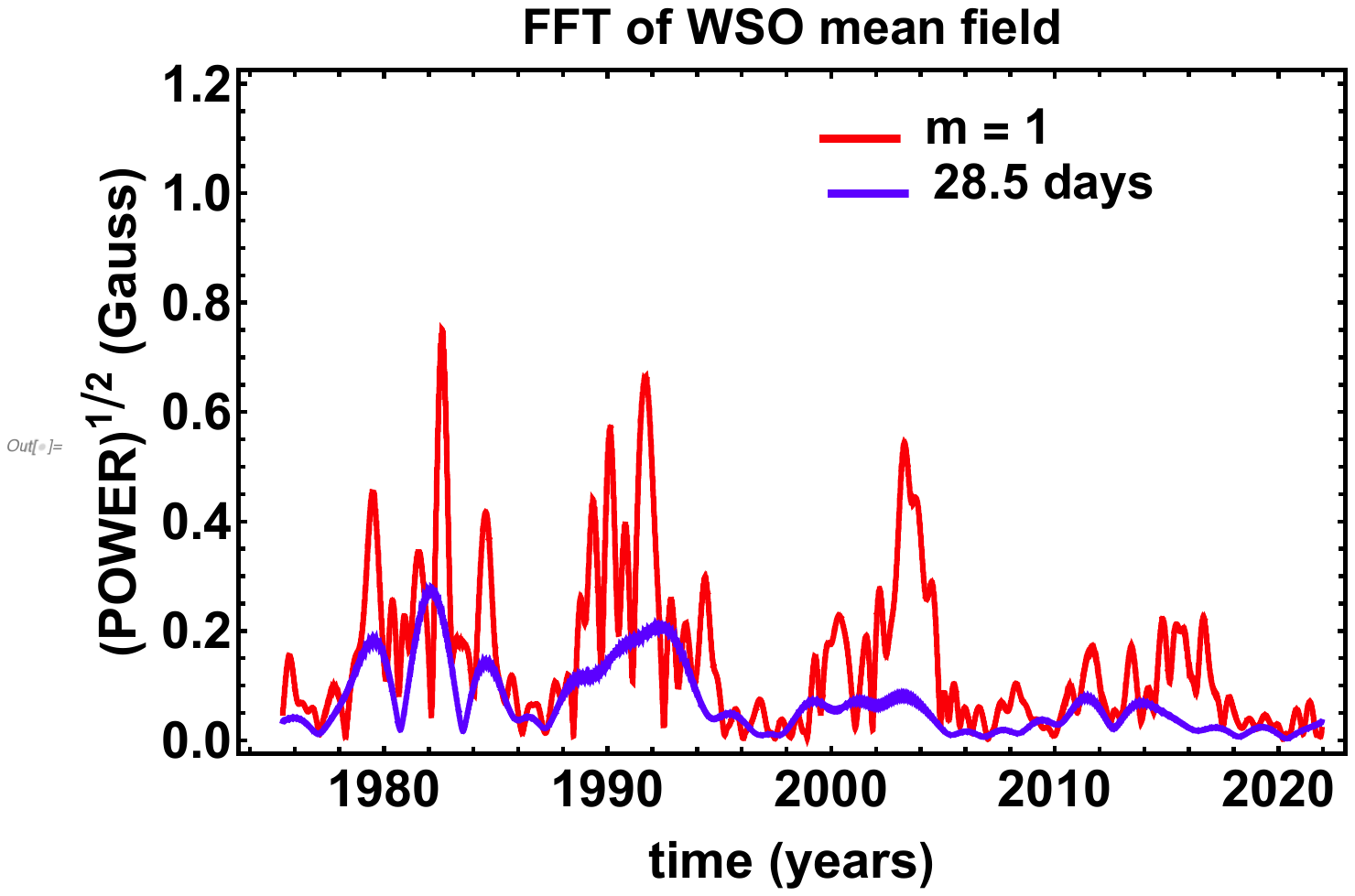}}}
 \centerline{
 \vspace{0.01in}
  \fbox{\includegraphics[bb=110 405 517 670,clip,width=0.60\textwidth]
 {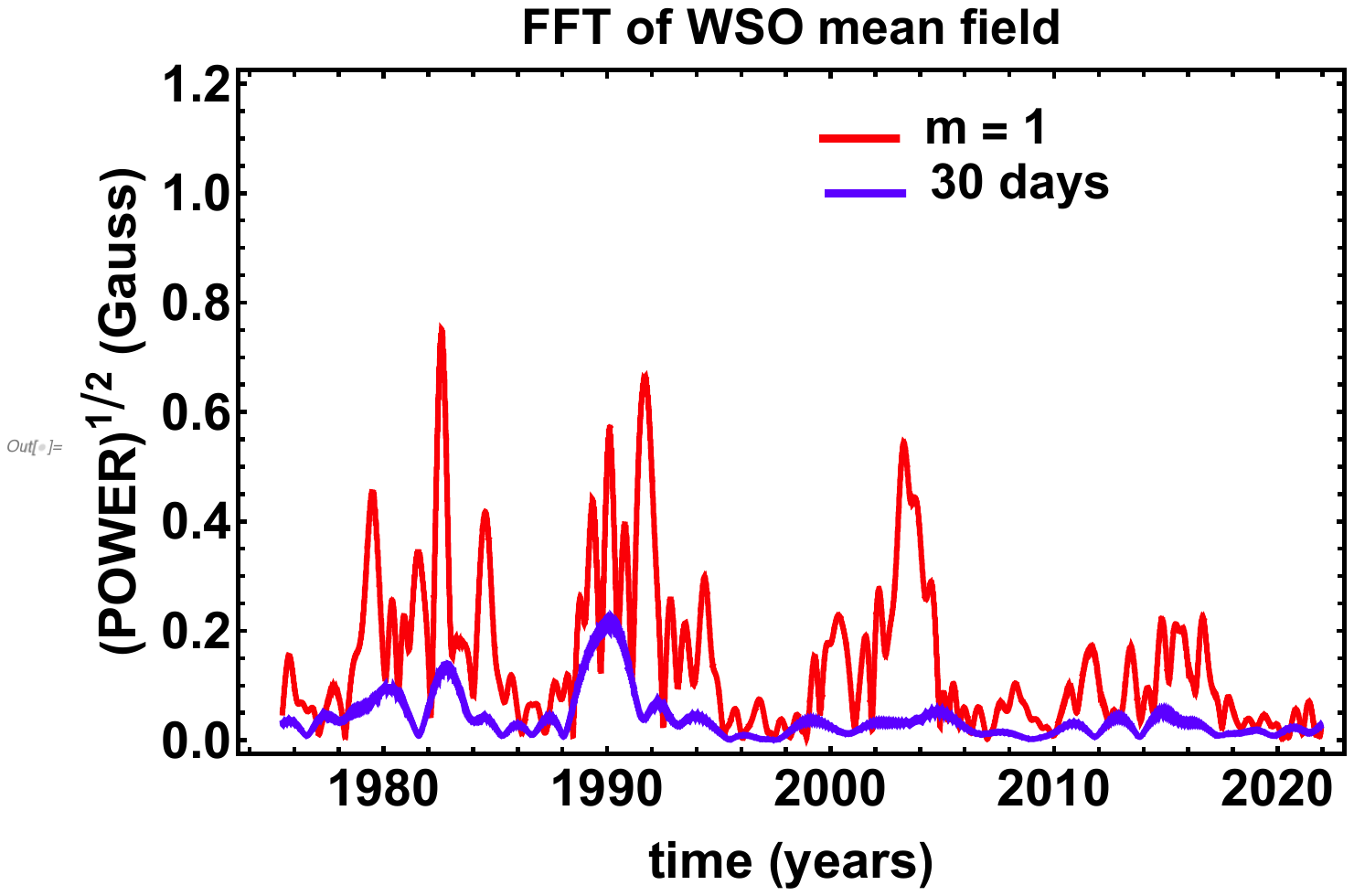}}}
\caption{ Two-sector ($m=1$) power from 27-day (top), 28.5-day (middle), and
30-day (bottom) structures (blue) compared with their total $m=1$
power (red).  This figure shows that 27-day power originates in all four sunspot cycles,
but the 28.5-day and 30-day power come mainly from the stronger sunspot cycles 21 and
22.
\label{fig:fig8}}
\end{figure}
\noindent
 and ${\omega} = 0.198-0.212~ \textrm{rad day}^{-1}$, we
obtain the blue curves in Figure~8.  Here, the red curve indicates the total $m=1$
power from the top panel of Figure~7.
So we are comparing the temporal origin of the individual $m=1$ peaks with the temporal
origin of the combined $m=1$ power.

Like Fourier transforms of continuous functions, these Fourier transforms of discrete
functions have peaks whose widths, ${\Delta}{\omega}$, are inversely related to the
lifetimes, ${\Delta}t$, of the corresponding temporal structures.  In fact,
${\Delta}{\omega}{\Delta}t~{\sim}~8$ for full widths at $e^{-1}$ maximum and
${\Delta}{\omega}{\Delta}t~{\sim}~8~ln2~{\approx}~5.54$ for full widths at half
maximum.  Consequently, the narrow $m=1$ peaks with
${\Delta}{\omega}~{\sim}~0.01~ \textrm{rad day}^{-1}$ in the left panel of
Figure~5 correspond to long-lived features with ${\Delta}t~{\sim}~1-2~ \textrm{yrs}$.

In the top panel of Figure~8 the blue curve refers to the power in the spectral `line'
at ${\omega} = 0.231~ \textrm{rad day}^{-1}$ (corresponding to a period of
approximately 27.2 days).  This blue curve tends to follow the
more rapidly fluctuating red curve, with appreciable contributions 
during each of the four sunspot cycles.  Thus, most of the two-sector power
originates in long-lived features that recur with a period of 27.2 days, and presumably
corresponds to quasi-vertical patterns in the 27.27-day Carrington stackplots of mean-field
observations.

In the middle panel, the blue curve refers to power in the spectral line at
${\omega}=0.219~ \textrm{rad day}^{-1}$ (28.7days).  Most of the 28.7-day power originates
in 1979-1985 and 1989-1993 coincident with large peaks of 27-day power.  This
overlapping trend did not continue into sunspot cycles 23 and 24 when the 28.7-day power
was much smaller.  This suggests that ${\sim}$28.5-day stackplot
patterns may have been weaker or less frequent in cycles 23 and 24 than in cycles
21 and 22.

In the bottom panel, the blue curve indicates power in the spectral line at
${\omega}=0.208~ \textrm{rad day}^{-1}$ (30.2 days).  Most of this $\sim$30-day
power occurred in 1989-1990, with lesser amounts during 1980 and 1982-1983, and
only trace amounts in sunspot cycles 23 and 24.

To summarize the results of Figure~8, the power depends on the rotation
period with 27-day power coming from all four sunspot cycles (but with a relatively
small contribution from the weakest sunspot cycle 24).  The 28.5-day power
originates mainly in sunspot cycles 21 and 22 with very small contributions from
cycles 23 and 24.  The 30-day power comes mainly from the year 1989
in cycle 22 and secondarily from small peaks in cycle 21.

The  lack of substantial 30-day power after 1990 provides another way to isolate the power
during 1989 - 1990.  We simply move the starting point of the Fourier transform backwards
in time through the year 1989 and watch the height of the 30-day peak increase.   Figure~9
shows a sample of the power spectra obtained by moving the starting time backward
from 02 February 1990 (CR1826) in steps of 3 Carrington rotations (approximately 82 days)
to 09 January 1989 (CR1811).
\begin{figure}[h!]
 \centerline{
 \fbox{\includegraphics[bb=110 334 528 660,clip,width=0.30\textwidth]
 {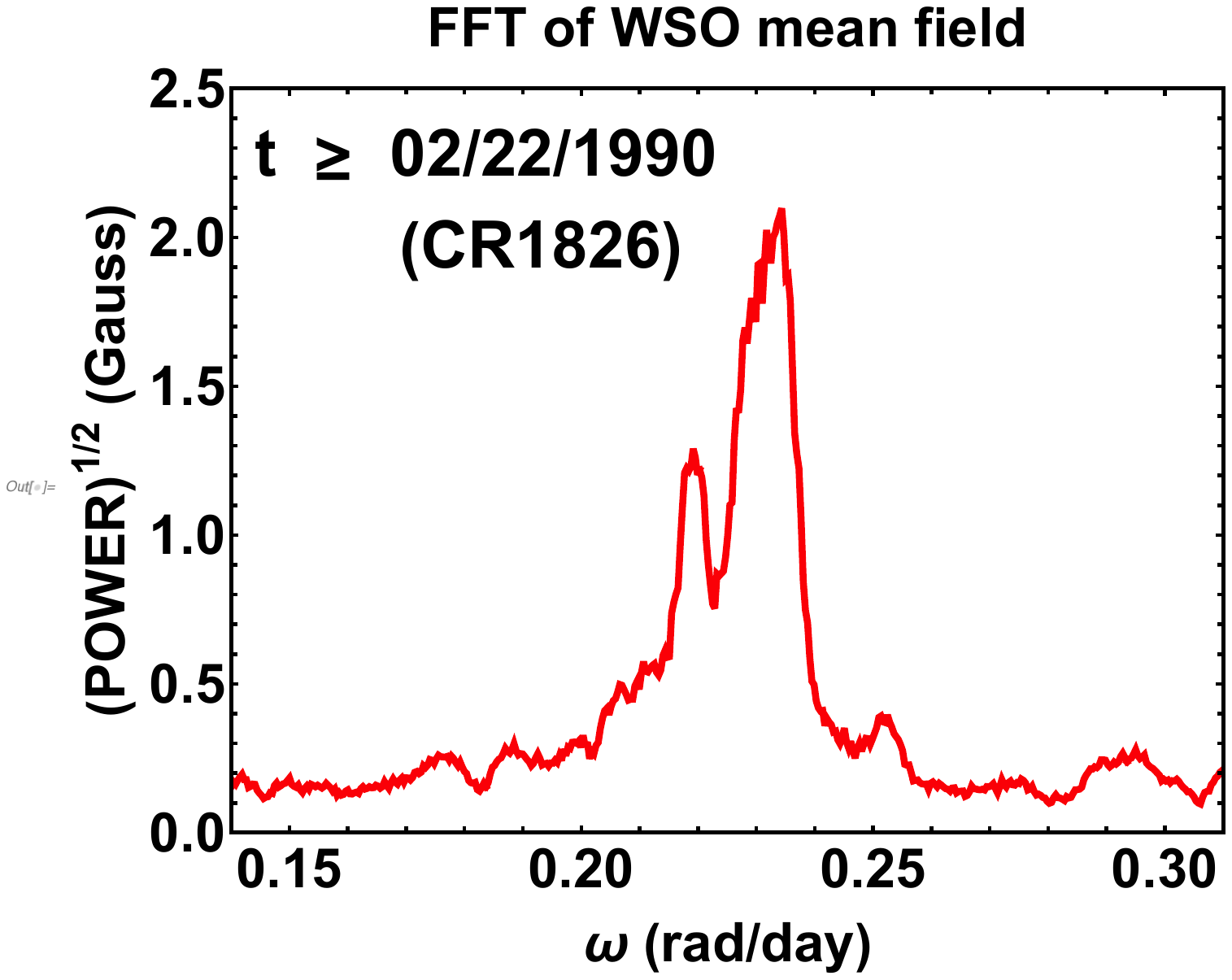}}
 \hspace{0.01in}
  \fbox{\includegraphics[bb=110 334 528 660,clip,width=0.30\textwidth]
 {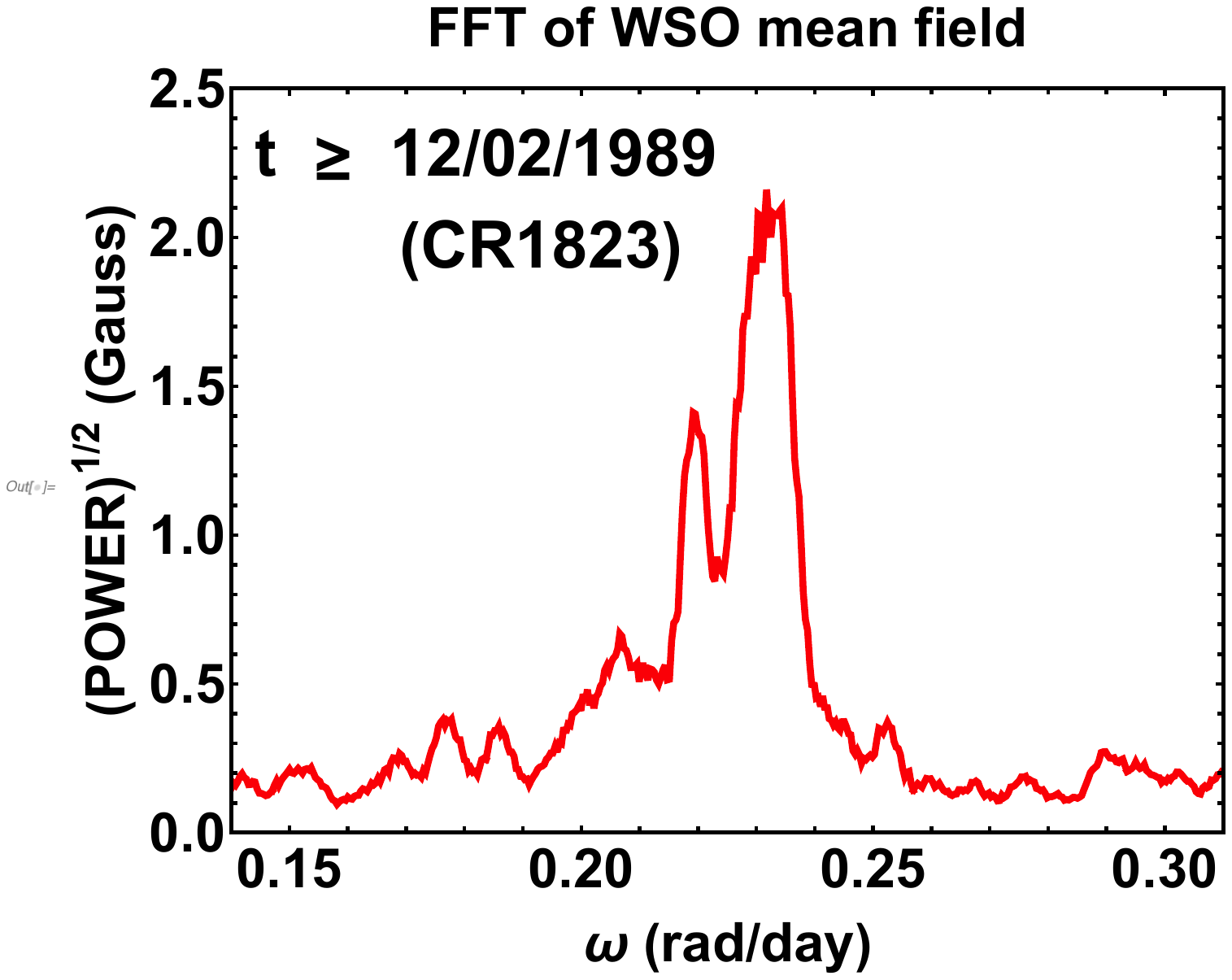}}
  \hspace{0.01in}
  \fbox{\includegraphics[bb=110 334 528 660,clip,width=0.30\textwidth]
 {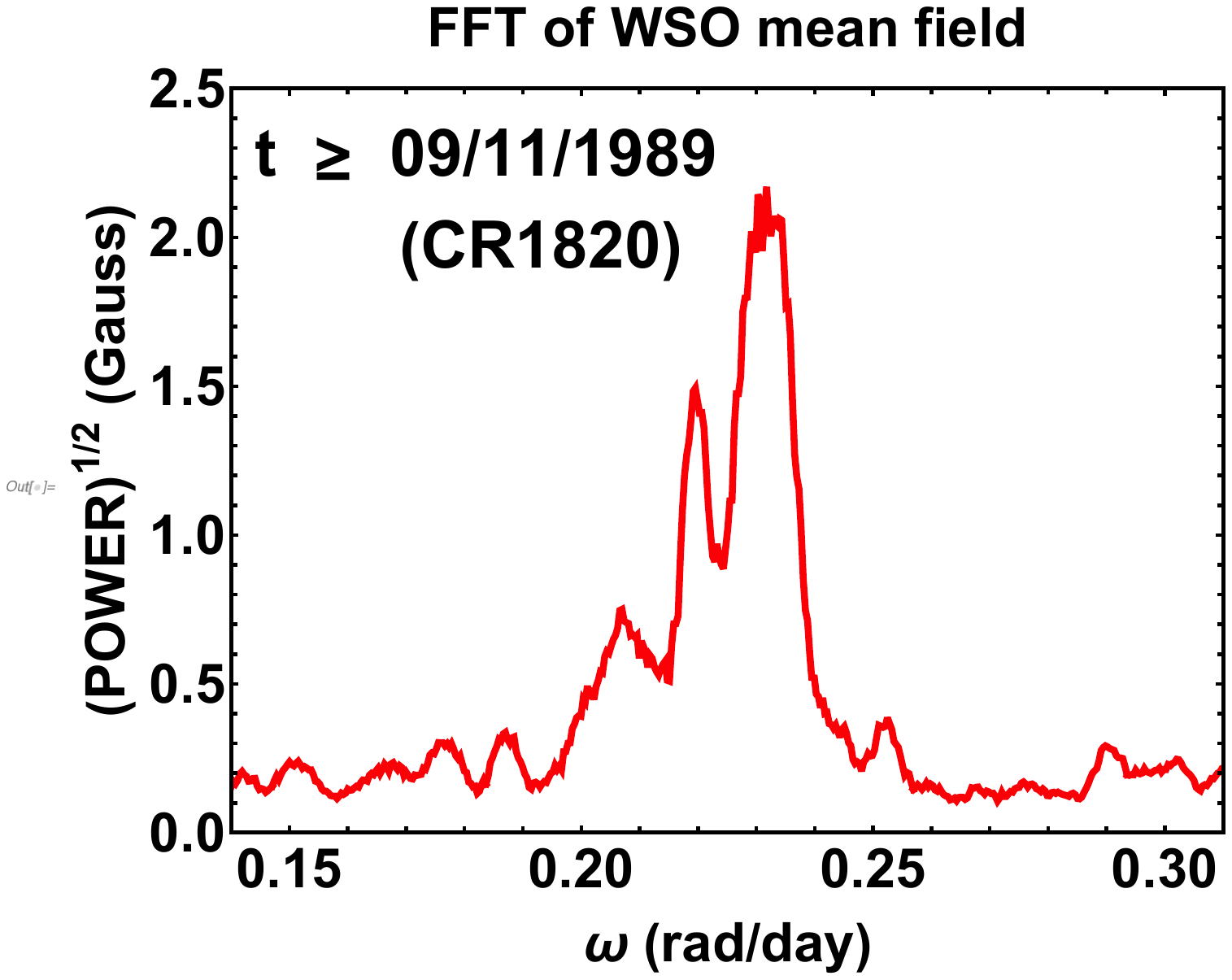}}}
  \centerline{
 \vspace{0.01in}
  \fbox{\includegraphics[bb=110 334 528 660,clip,width=0.30\textwidth]
 {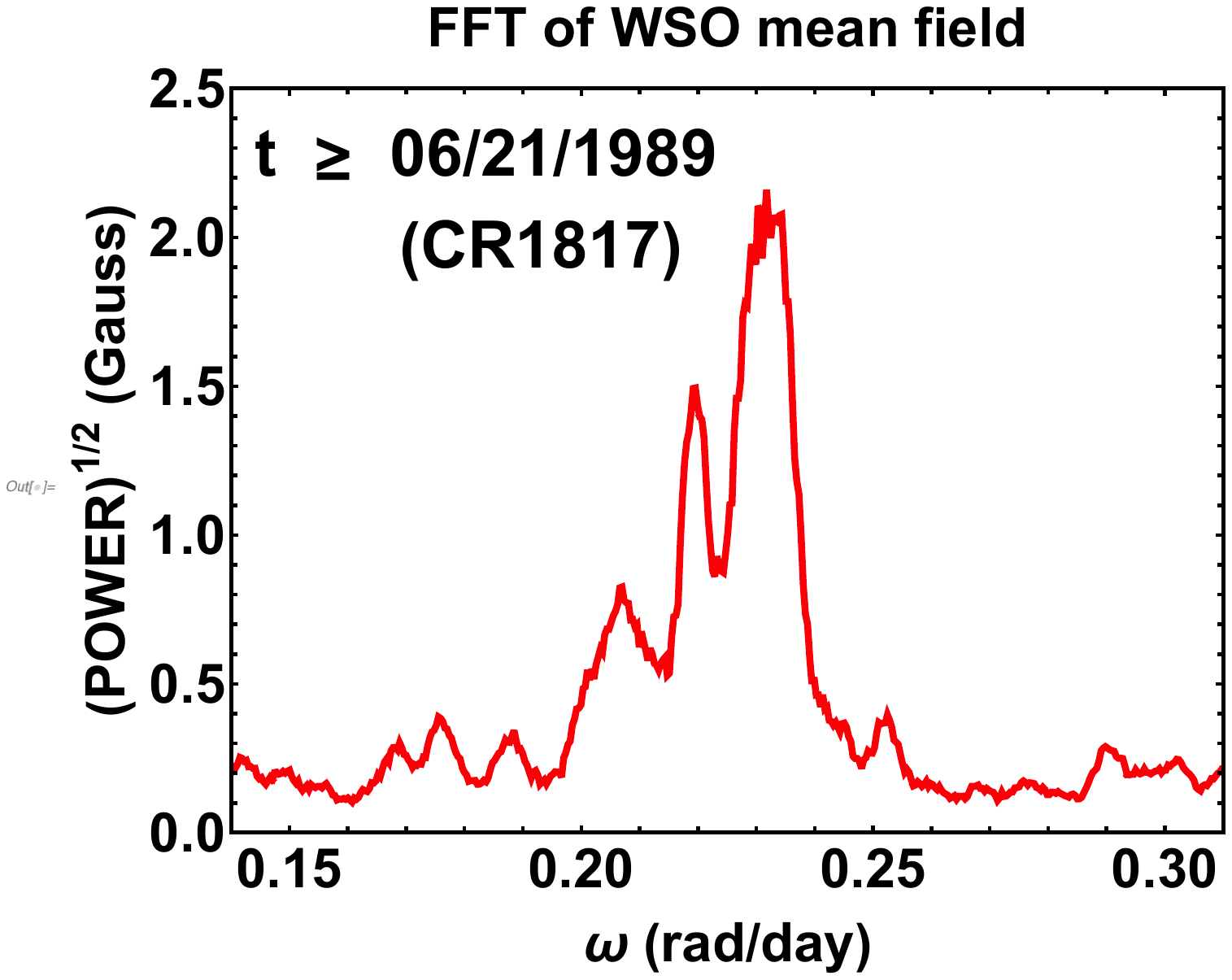}}
  \hspace{0.01in}
  \fbox{\includegraphics[bb=110 334 528 660,clip,width=0.30\textwidth]
 {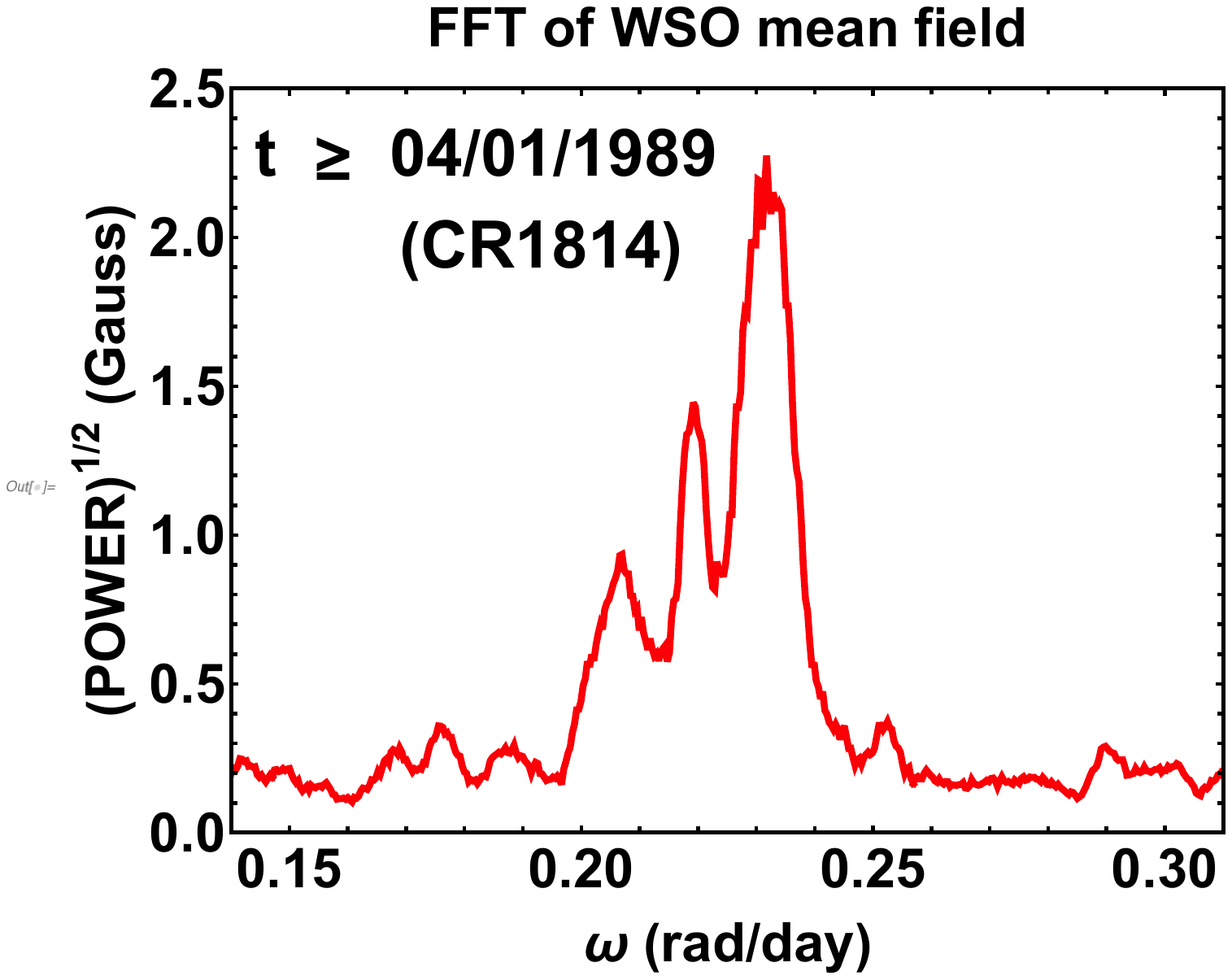}}
  \hspace{0.01in}
  \fbox{\includegraphics[bb=110 334 528 660,clip,width=0.30\textwidth]
 {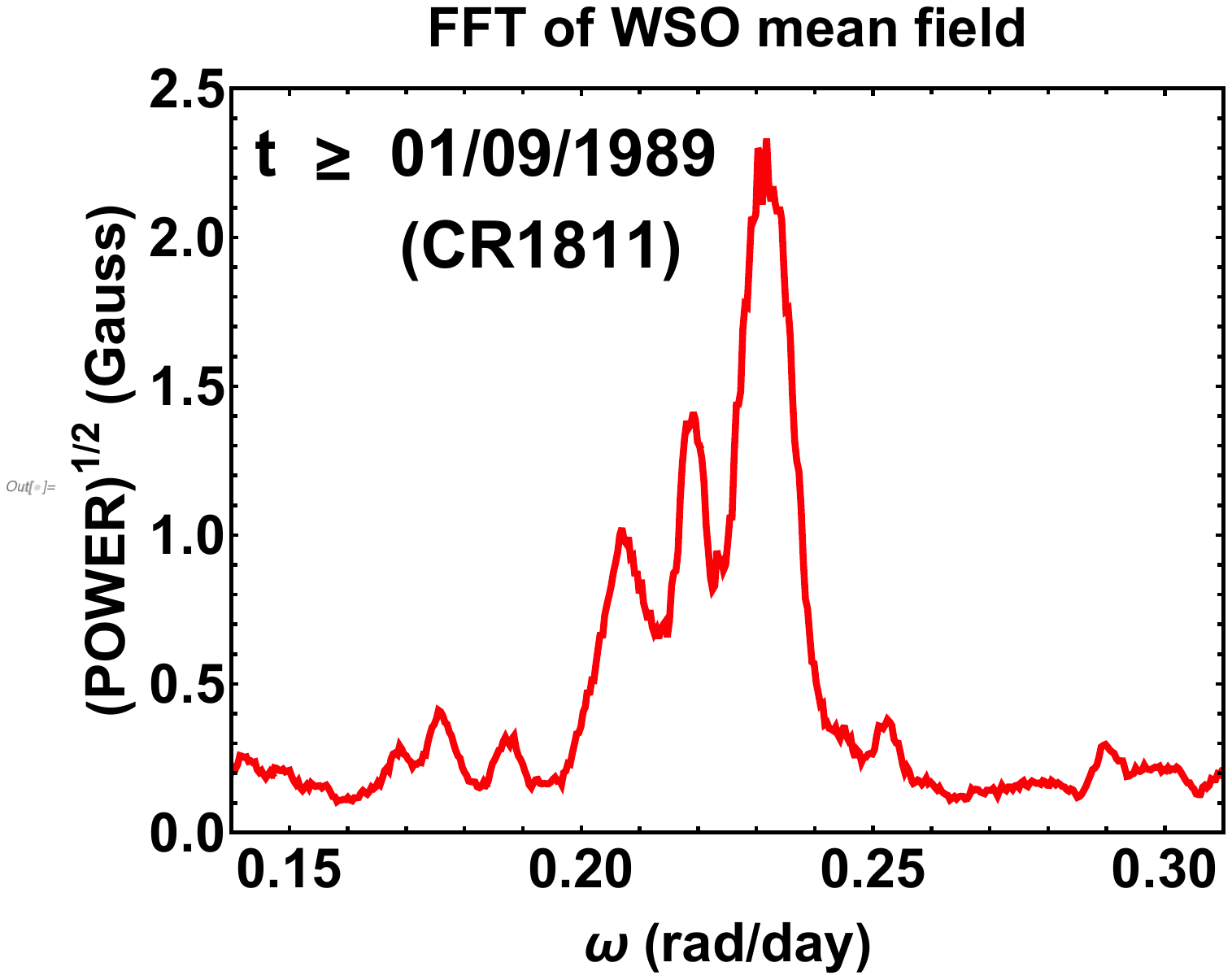}}}
\caption{Power spectra of WSO mean-field measurements from time, t,
to the end of the data set on 16 November 2021,
showing the emergence of the ${\sim}$30-day
($0.208~ \textrm{rad day}^{-1}$) peak as t decreases
from 22 February 1990 to 09 January 1989.
\label{fig:fig9}}
\end{figure}
During this sequence, the 30-day ($0.208~ \textrm{rad day}^{-1}$) peak
emerges from a continuum level at about 0.5 G to a maximum height of about
1.0 G.  Although not shown here, a movie with 1-rotation time resolution indicated
that the 30-day peak emerged from the continuum at CR1824 (29 December 1989)
and strengthened until it reached about 1.0 G at CR1811 (09 January 1989),
corresponding to a lifetime of 13 rotations (1 yr and 11 days).  Thus, the 30-day
oscillation spanned the 1-year interval from 1989 to 1990.

We can continue this approach by moving the starting point of the Fourier transform
forward in time to
successively exclude major contributions to the 30-day, 28.5-day, and 27-day
periods.  Referring to Figure~8, we select the first starting time on
09 January 1989 when substantial power remained in
all three rotational periods.  This is shown in the upper-left panel of
Figure~10.  Then, we move farther in time to 01 June 1991, which is well after
the peak of 30-day power, but still includes power at 28.5 days and 27-days,
as shown in the upper-right panel of Figure~10.  Next, we choose 17 January 1996,
which is after the large peak of 28.5-day power.  However, 17 January 1996
is still before the occurrence of the large peak of 27-day power in 2003-2004,
and this contribution is shown in the lower-left panel of Figure~10.  Finally,
we select 02 January 2005 to remove this large peak of 27-day power.  The
remaining 27-day power comes from a small peak in 2015 -2016, as shown in
the lower-right panel of Figure~10.  This contribution seemed very important
when it rejuvenated the large-scale field in sunspot cycle 24
\citep{2015ApJ...809..113S}.  In the next section, we shall compare these
observations with spatially resolved magnetograms and find that the peaks
of 27-day power tend to occur toward the end of sunspot maximum when the
sunspot belts are closer together and allow large unipolar magnetic regions to
form at the equator. 
\begin{figure}[h!]
 \centerline{
 \fbox{\includegraphics[bb=110 335 530 665,clip,width=0.30\textwidth]
 {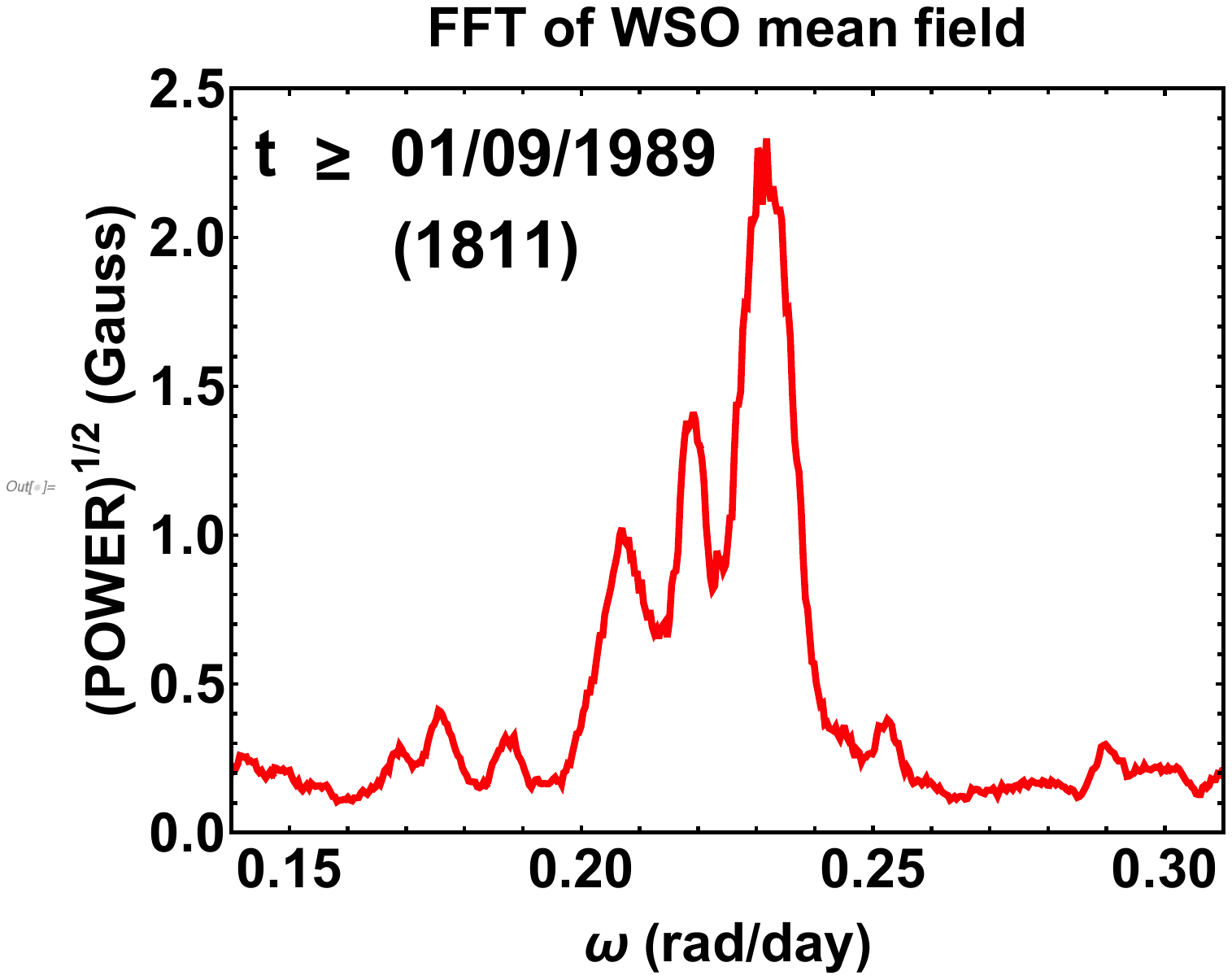}}
 \hspace{0.01in}
  \fbox{\includegraphics[bb=110 335 530 665,clip,width=0.30\textwidth]
 {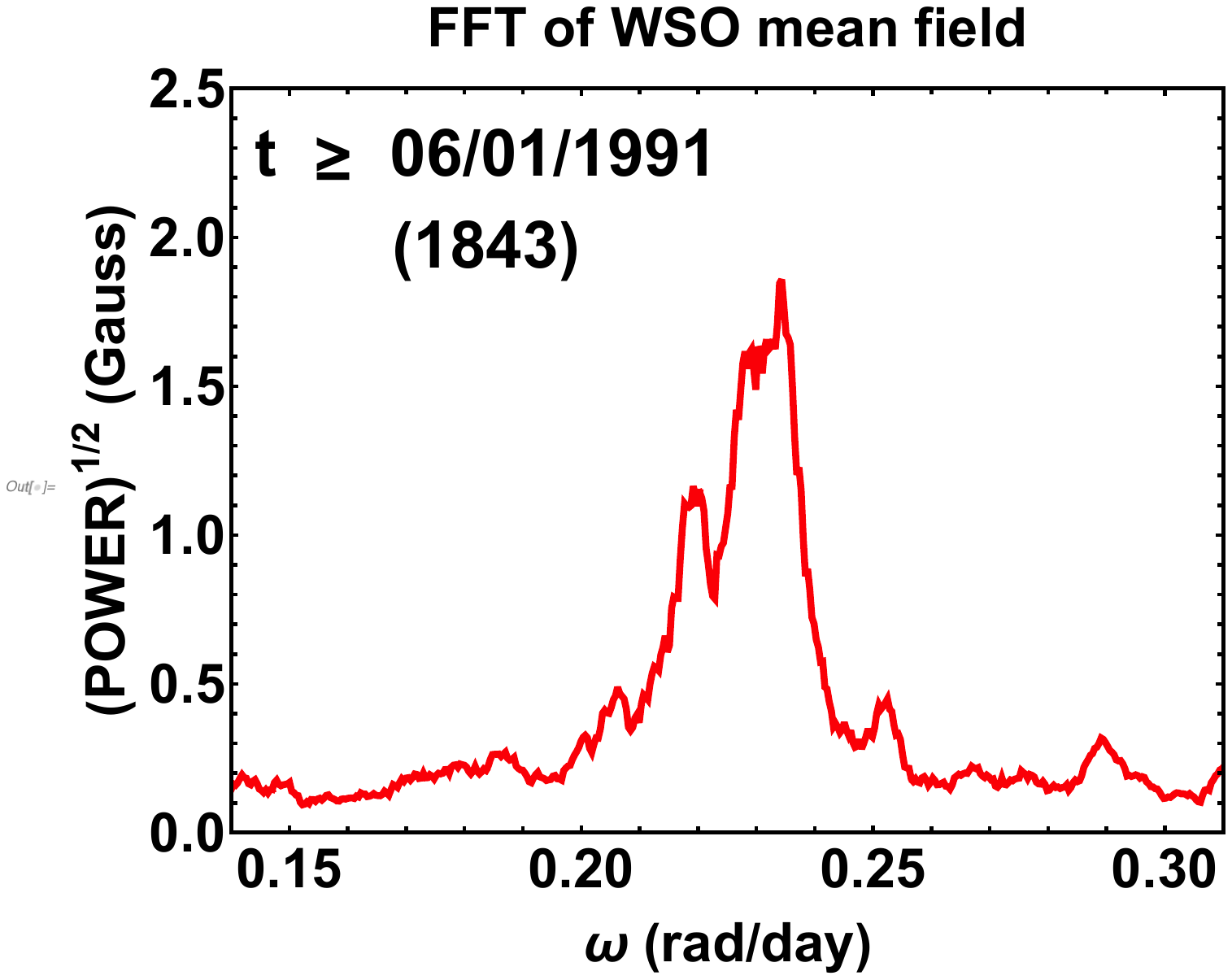}}}
  \centerline{
 \vspace{0.01in}
  \fbox{\includegraphics[bb=110 335 530 665,clip,width=0.30\textwidth]
 {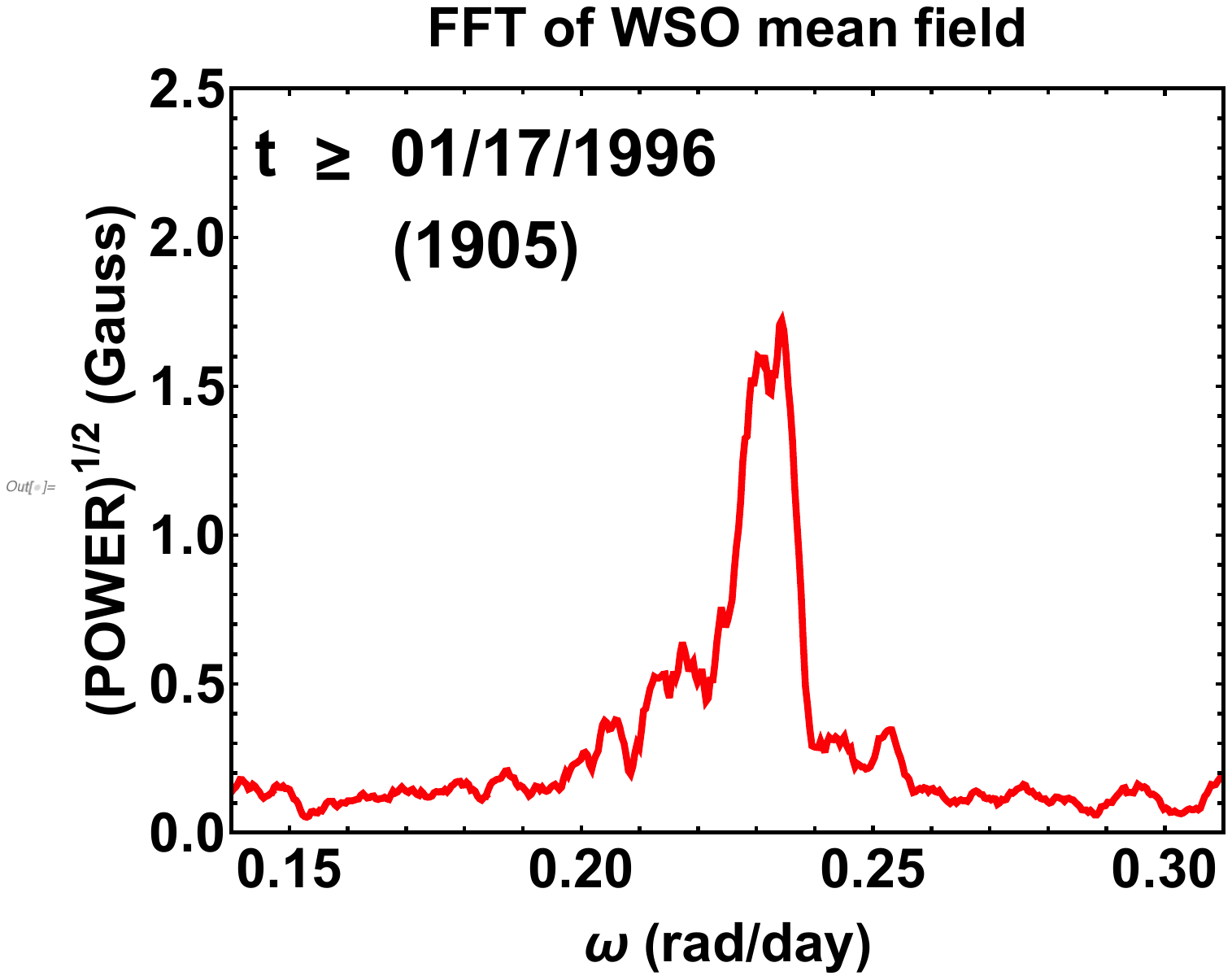}}
  \hspace{0.01in}
  \fbox{\includegraphics[bb=110 335 530 665,clip,width=0.30\textwidth]
 {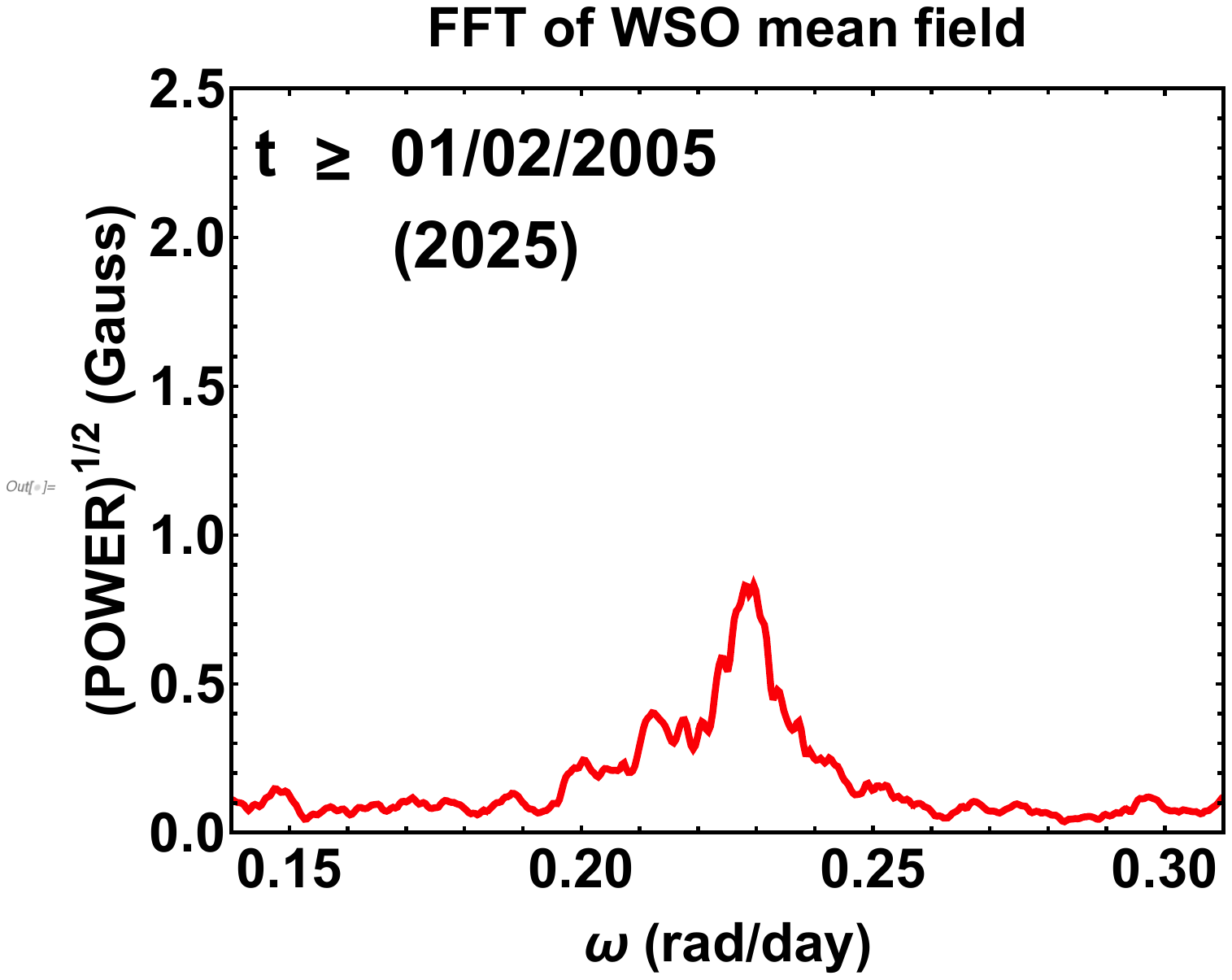}}} 
\caption{Power spectra of the mean field, showing the $m=1$
fine structures at 0.21, 0.22 and $0.23~ \textrm{rad day}^{-1}$ as the starting
time, t, of the Fourier transform is increased.
The 30-day and 28.5-day sidebands are successively reduced
as t shifts from 1989 to 1991 and then to 1996.  The 27-day peak is reduced
when t shifts to 2005, leaving a small contribution from 2015 - 2016.
\label{fig:fig10}}
\end{figure}

\subsubsection{Corresponding solar images}
Next, we look for these
 magnetic patterns in spatially resolved solar observations.
We begin with Figure~11, which shows Carrington maps
of the photospheric magnetic field obtained at the National Solar
Observatory (NSO)\footnote{https://nispdata.nso.edu/ftp/kpvt/synoptic/mag}.
At NSO, each of these
maps was divided by ${\mu}=\sin{\theta}\cos{\phi}$ to convert the observed
line-of-sight component
to a radial component, assuming that the fields are radial at the photosphere
where they are measured.  Thus, we regard these maps as displays of the radial
component of photospheric field.
\begin{figure}[h!]
 \centerline{
\fbox{ \includegraphics[bb=20 05 595 790,clip,angle=-90,width=0.9\textwidth]
 {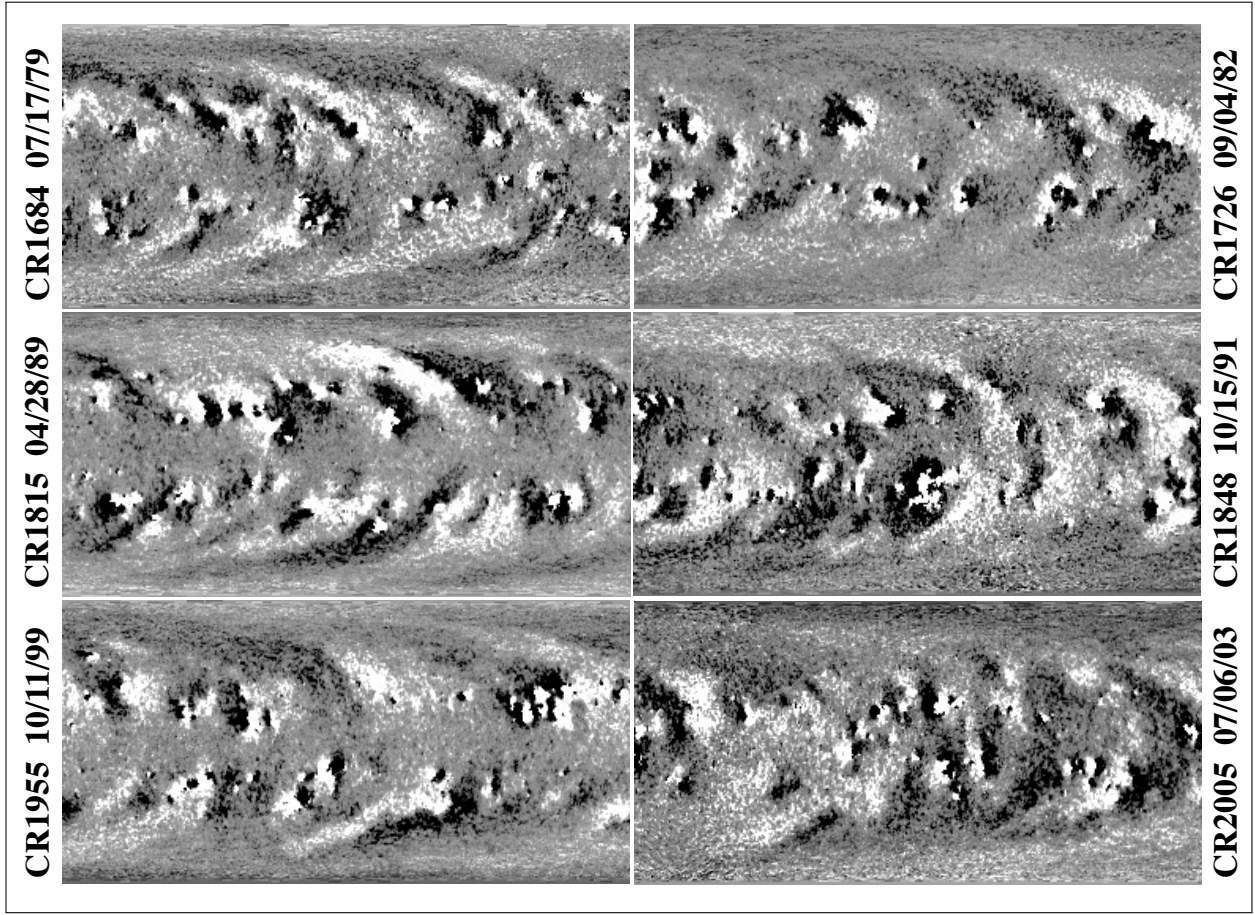}}}
\caption{NSO Carrington maps of the radial magnetic field during sunspot cycles
21 - 23 (top, middle, and bottom panels, respectively), showing flux streaming
poleward and eastward from the sunspot belts during the initial phase of sunspot
maximum (left panels) and flux accumulating in the equatorial region between the
sunspot belts at the end phase of sunspot maximum (right panels).  Longitude runs
left to right from 0$^{\circ}$ to 360$^{\circ}$, and sine latitude ranges from -1 to +1
from bottom to top of each map.   Positive polarity is white and negative is black.
Dates refer to the starting times at the right edge of each map. 
\label{fig:fig11}}
\end{figure}

This figure contains images from the start of sunspot maximum (left column)
and the end of sunspot maximum (right column) in sunspot cycles 21 (top row),
22 (middle row), and 23 (bottom row).  In the left panels,  each sunspot cycle has
progressed far enough that several large active regions have emerged to form
activity belts with flux streaming poleward and eastward (leftward) from those belts.
However, the sunspot cycles have not progressed far enough for flux to spread
equatorward to fill in the wide gaps between the two belts, as has happened
at the times of the images in the right panels.  In fact, by the end of the
sunspot-maximum era, the two activity belts have reached sufficiently low latitudes
and narrow separations that the equatorward diffusion of flux dominates the poleward
convection by meridional flow, causing flux to accumulate around the equator.

We have encountered these phenomena before.  First, in numerical simulations of
the mean field, \cite{1986SoPh..104..425S} found that the 28-29-day
recurrent patterns originated in flux that was migrating poleward from its
sources in the sunspot belts.  The dramatic eastward drift of these patterns is well
known to many of us from viewing time-lapse movies of Carrington maps like the
ones in Figure~11.  Second, \cite{2015ApJ...809..113S} found that a
juxtaposition of northern-hemisphere and southern-hemisphere active regions
during the second half of 2014 created a large region of positive-polarity flux at the
equator, which produced a major rejuvenation of the Sun's large scale field.  As shown
in the top panel of Figure~8, the 27-day power in the two-sector component of the
mean field reached a peak at this time.  Also, \cite{2015ApJ...809..113S} noted
that this rejuvenation of the large-scale field was not an isolated characteristic of
sunspot cycle 24, and that similar enhancements of the equatorial dipole field in
1982, 1991, and 2003 have marked the end of the sunspot maximum era (or the
start of the declining phase) of cycles 21, 22, and 23.

Another reason for selecting the images in Figure~11 was to relate these flux
distributions to the profiles of spectral power, especially for the two-sector
($m=1$) plots in Figure~8.  Thus, the maps in the top panel of Figure~11
occur in 1979 and 1982 when there were peaks in the spectra
for 27 days, 28.5 days and, to a lesser extent, 30 days.  The maps in the middle
panels occurred in 1989 when the 30-day power reached its maximum,
and in 1991 when the 27-day and 28.5-day power reached their maximum values.
The map in the bottom-right panel was chosen because it occurred in 2003
when the 27-day power dominated the two-sector spectrum.   As one can see, it shows
a large two-sector pattern of equatorial flux with positive-polarity flux left of center and
negative-polarity flux right of center.

In choosing these images, I looked for strong poleward streams in each sunspot cycle.
However, I did not always find them.   Whereas the maps in the upper-left and middle-left panels show major streams, the map in the lower-left panel (CR1955, 11 October 1999)
shows relatively weak streams, despite the fact that it was at the same phase of the
sunspot cycle.  Those 1999 streams are only slightly more impressive than the weak
streams in 1982, 1991, and 2003 in the right column.  This favoring of sunspot
cycles 21 and 22 over cycle 23 is consistent with the power spectra in Figure~8,
which show more 28.5-day and 30-day power during cycles 21 and 22 than during
cycle 23.

If the 28.5-day power originates in poleward migrating flux from large active
regions, as previously reported \citep{1986SoPh..104..425S}, then it seems
plausible that the much rarer 30-day power in 1989 is a statistical fluctuation caused
by the emergence of an especially large, high-latitude active region at that time.  The
strong, northern-hemisphere stream in the middle-left panel of Figure~11 originated
in such an active region.

The right panel of Figure~12 shows the evolution of this region during CR1811-1818
(09 January 1989 - 19 July 1989) when sunspot cycle 22 reached the start of its
3-4 years of high sunspot activity\footnote{J.W. Harvey recently reminded me that
during CR1813, this region (5395) was the source of many X-ray flares and coronal
mass ejections (CMEs), two of which were responsible for the blackout of the
Hydro-Qu\'ebec power system on 13 March 1989 \citep{2019SpWea..17.1427B}.}.
Each image is the northern-hemisphere
part of a Carrington map that has been cropped at the equator.  Thus, longitude
runs left to right from 0$^{\circ}$ to 360$^{\circ}$ and sine latitude runs bottom
to top from 0 to +1.  The faint, yellow lines provide a reference drift for a
30.2-day rotation, corresponding to the frequency
${\omega}$ = 0.208 $\textrm{rad~day}^{-1}$ that we obtained from the power
spectra in Figures~4 and 5.

For comparison, the left panel of Figure~12 shows the evolution of
a much smaller region during CR1788-1795 (22 April - 30 October 1987) at the
\begin{figure}[h!]
 \centerline{
\fbox{ \includegraphics[bb=15 071 600 735,clip,angle=-90,width=0.85\textwidth]
 {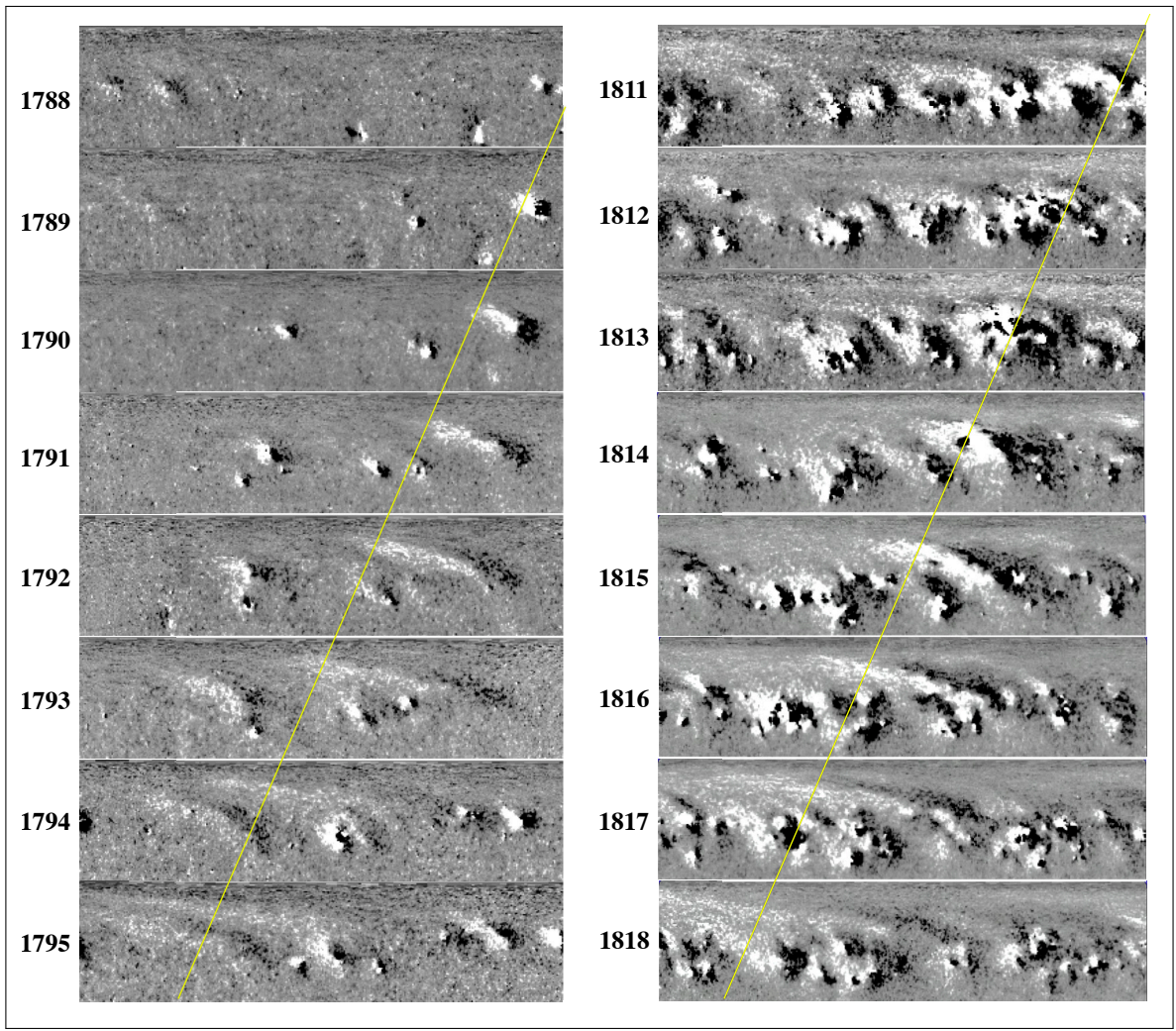}}}
\caption{Carrington maps, cropped to show northern-hemisphere magnetic fields
as a function of longitude (0$^{\circ}$ to 360$^{\circ}$ on the horizontal axis) and
sine latitude (0 to +1 on the vertical axis).  (left):  CR1788-1795 (22 April - 
 30 October 1987), showing the northward and eastward drift of flux from a new-cycle
 active region.  (right): CR1811-1818 (09 January - 19 July 1989), showing a
stronger stream from a bigger active region near sunspot maximum.  Faint yellow
lines indicate a 30.2-day rotation.
\label{fig:fig12}}
\end{figure}
\noindent
start of cycle 22.  Both regions emerged at latitudes ${\sim}$35$^{\circ}$, and their
fluxes evolved into similar patterns at comparable speeds.  (An interesting, and probably coincidental, similarity is that both streams were replenished by flux from a second
active region that emerged 3-4 rotations later toward the left side of each panel.)
The major difference is that the 1989 source was larger and stronger than the
1987 source.  Presumably, this difference in strength was responsible for the
difference in spectral power shown in Figure~8.  The 1989 stream coincided with a
large peak of 30-day spectral power, whereas the 1987 stream did not.

At the ${\sim}$35$^{\circ}$ latitude of these two active regions, the
rotation period of small-scale magnetic tracers is about 29 days
\citep{1983ApJ...270..288S}.  However, within a few Carrington rotations, these streams
of trailing, positive-polarity flux had drifted poleward into the 45$^{\circ}$ latitude range
where the rotation period is 30 days.  Although one can confirm this by measuring the
vertical locations of the streams in these images, one can also refer to the faint yellow
lines, which provide a 30.2-day reference drift.  These lines track the trailing streams of positive-polarity flux fairly well until the tails of the streams reach latitudes above
45$^{\circ}$ and also begin to merge with flux from other active regions.

Multi-latitude stackplots of the WSO-derived source-surface
field support the identification of the 1989 pattern as the source of the 30-day
power \citep{1994JGR....99.6597W}.
Calculated for 1977 - 1993, these stackplots show a rare, 30-day, two-sector
recurrence pattern over a wide range of latitudes from 20$^{\circ}$N to
80$^{\circ}$N during 1989.  It is the most prominent 30-day pattern during that
16-year interval.  In addition, its positive-polarity sector moves from right to left across
the Carrington frame during 1989, so that the phase of this two-sector pattern
also agrees with that of the field in Figure~12.

The `1989 pattern' continued its poleward and eastward migration well after the end of
1989 when the 30-day power ended, according to Figure~9.  By CR1830 (11 June 1990),
the trailing end of the
stream extended to about  68$^{\circ}$ latitude where the \cite{1983ApJ...270..288S}
rotation period is 35.3 days.  However, the rotation period of the mean field did not
increase beyond 30 days.  Thus, for latitudes above 45$^{\circ}$, even this relatively
strong field was too weak to overcome the ${\mu}^{2}$-dependence of the mean-field
integral in Eq(3), and (as we shall see in the Appendix), an extra factor of $\mu$ due to
limb darkening.  In the next section, we shall see that this result provides a clue for
understanding some puzzling associations between the axisymmetric component of the
mean field and the Sun's polar magnetic fields.

\subsubsection{Power in the Axisymmetric Component of the Sun's Mean Field}
To display the non-axisymmetric part of the mean field, $B_{m}$, we used the
27-day running mean of the standard deviation, defined by $B_{m}^{rms}=(<B_{m}^2>-<B_{m}>^2)^{1/2}$,
where the brackets refer to averages over a 27-day moving window.  In this subsection,
we are interested in $<B_{m}>$, the 27-day moving average of the mean field (\textit{i.e.} the
axisymmetric component that we squared  and subtracted from $<B_{m}^{2}>$ to get that
non-axisymmetric component).

Another way to compare the axisymmetric and non-axisymmetric components of $B_{m}$
is to express Eq(7) in terms of real variables and separate the $m=0$ and $m{\neq}0$ terms.
To first order in $B_{0}$, we obtain
\begin{equation}
B_{m}~{\approx}~\left \{ \sum_{l=1}^{\infty} {\rho}_{l0} I_{l0}~+~2 \sum_{l=1}^{\infty} \sum_{m=1}^{l}
 {\rho}_{lm} I_{lm} \cos{\delta_{lm}} \right \} ~+~B_{0} \left \{\sum_{l=1}^{\infty} {\rho}_{l0} J_{l0}~+~2 \sum_{l=1}^{\infty} \sum_{m=1}^{l} {\rho}_{lm} J_{lm} \cos{\delta_{lm}} \right \}.
\end{equation}
Averaging over ${\delta}_{lm}$ (which we assume varies linearly with time, $t$, or more precisely
with $mt$), we get
\begin{equation}
B_{m}^{rms}~{\approx}~\sqrt{2}~\sqrt{ {\rho^{2}_{11}} I_{11}^2~+~{\rho^{2}_{22}} I_{22}^2~+~
{\rho^{2}_{33}} I_{33}^2~~+ {\rho^{2}_{31}} I_{31}^2   }
\end{equation}
for the root-mean-square deviation of $B_{m}$ from its average value, and
\begin{equation}
<B_{m}>~{\approx}~{\rho}_{20}I_{20}~+~B_{0} \left \{{\rho}_{10} J_{10}~+~
{\rho}_{30}J_{30} \right \}
\end{equation}
for the 27-day moving average of $B_{m}$.  In Eq(16), we have omitted
$J_{lm}$-dependent terms because they occur with the second-order factor,
$B_{0}^2$, and we have omitted terms with $I_{21}$ and $I_{32}$, which vanish,
as indicated in Table~1.  Likewise, in Eq(17), we have omitted terms with $I_{10}$,
$I_{30}$, and $J_{20}$, which also vanish as indicated in Tables~1 and 2.  Thus,
$B_{m}^{rms}$ has non-axisymmetric ($m~{\neq}~0$) contributions from $Y_{1}^{1}$,
$Y_{2}^{2}$, $Y_{3}^{3}$, and $Y_{3}^{1}$ for which $I_{11}=+0.173$, $I_{22}=+0.103$,
$I_{33}=+0.035$, and $I_{31}=-0.027$, as obtained from Table~1.  In that case,
we expect the
non-axisymmetric component of $B_{m}$ to be dominated by contributions from $Y_{1}^{1}$
and $Y_{2}^{2}$ with much smaller contributions from $Y_{3}^{3}$ and $Y_{3}^{1}$.

The axisymmetric quantity, $<B_{m}>$, has contributions from $Y_{2}^{0}$ with
$I_{20}=-0.084$, plus $B_{0}$-dependent contributions from $Y_{1}^{0}$ and $Y_{3}^{0}$,
with the somewhat larger values of $J_{10}=+0.244$ and $J_{30}=-0.093$.  However,
the factor of $B_{0}$ reduces these contributions and modulates them with a 1-year period.
When $B_{0}$ has its maximum value of $7.25^{\circ}$ (0.1265 rad),
$B_{0}J_{10}$ is only 0.031 (37\% of $I_{20}$) and $B_{0}J_{30}$ has the even smaller
value of -0.0118 (38\% of $B_{0}J_{10}$).
Thus, in rough terms, we can regard $Y_{2}^{0}$, $Y_{1}^{0}$, and $Y_{3}^{0}$ as making
three monotonically decreasing contributions to the axisymmetric part of $B_{m}$, in which each contribution is about 40\% of the previous one.  Now, let us see how well these terms fit the WSO mean-field measurements. 
  
The top panel of Figure~13 contains a plot of $<B_{m}>$ (red) and the monthly averaged
sunspot number from the Royal Observatory of Belgium (SILSO) (blue) during sunspot
cycles 21 - 24.  The sunspot number has been divided
\begin{figure}[h!]
 \centerline{
 \fbox{\includegraphics[bb=110 420 500 673,clip,width=0.61\textwidth]
 {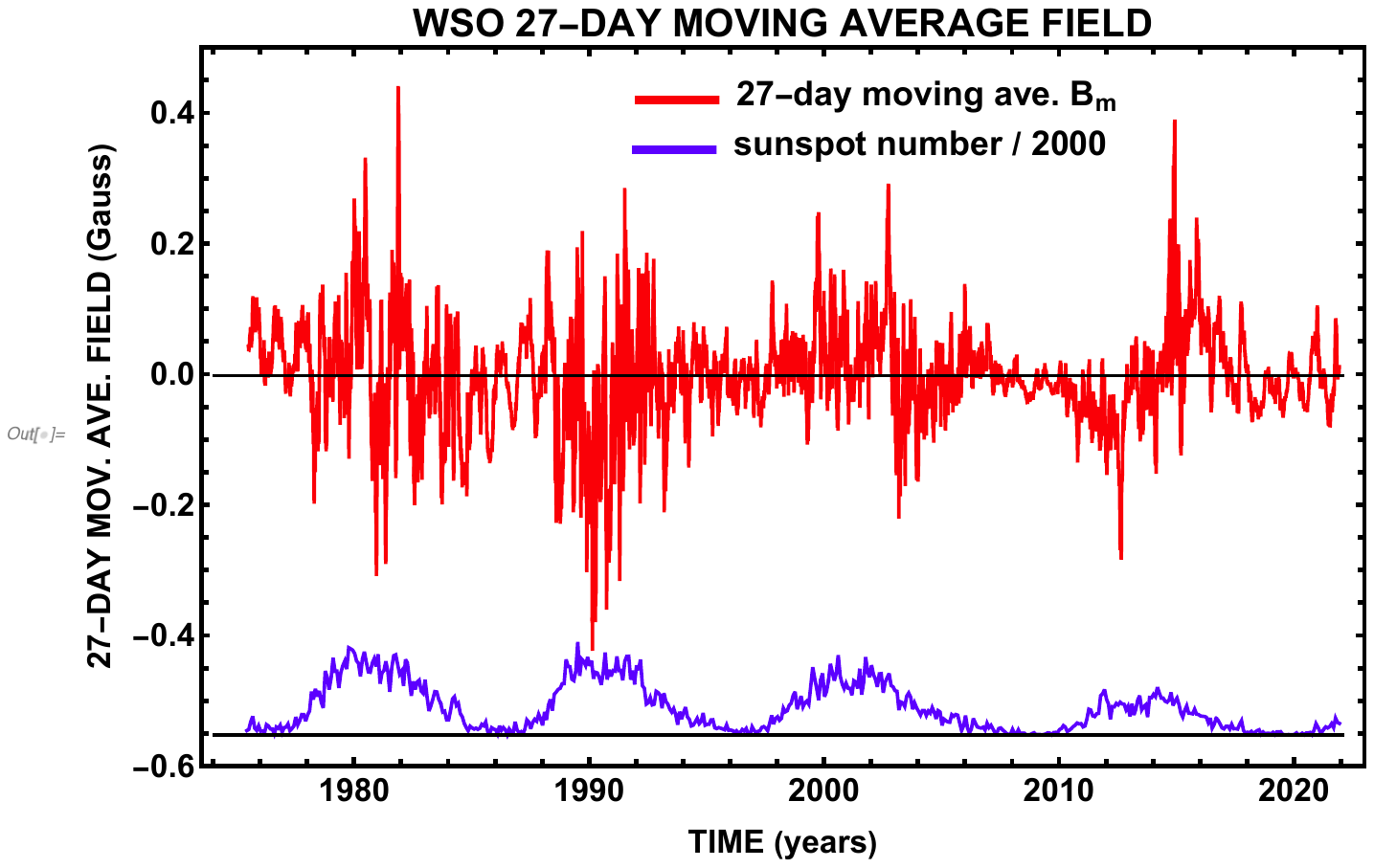}}}
  \centerline{
 \vspace{0.01in}
  \fbox{\includegraphics[bb=110 420 500 673,clip,width=0.61\textwidth]
 {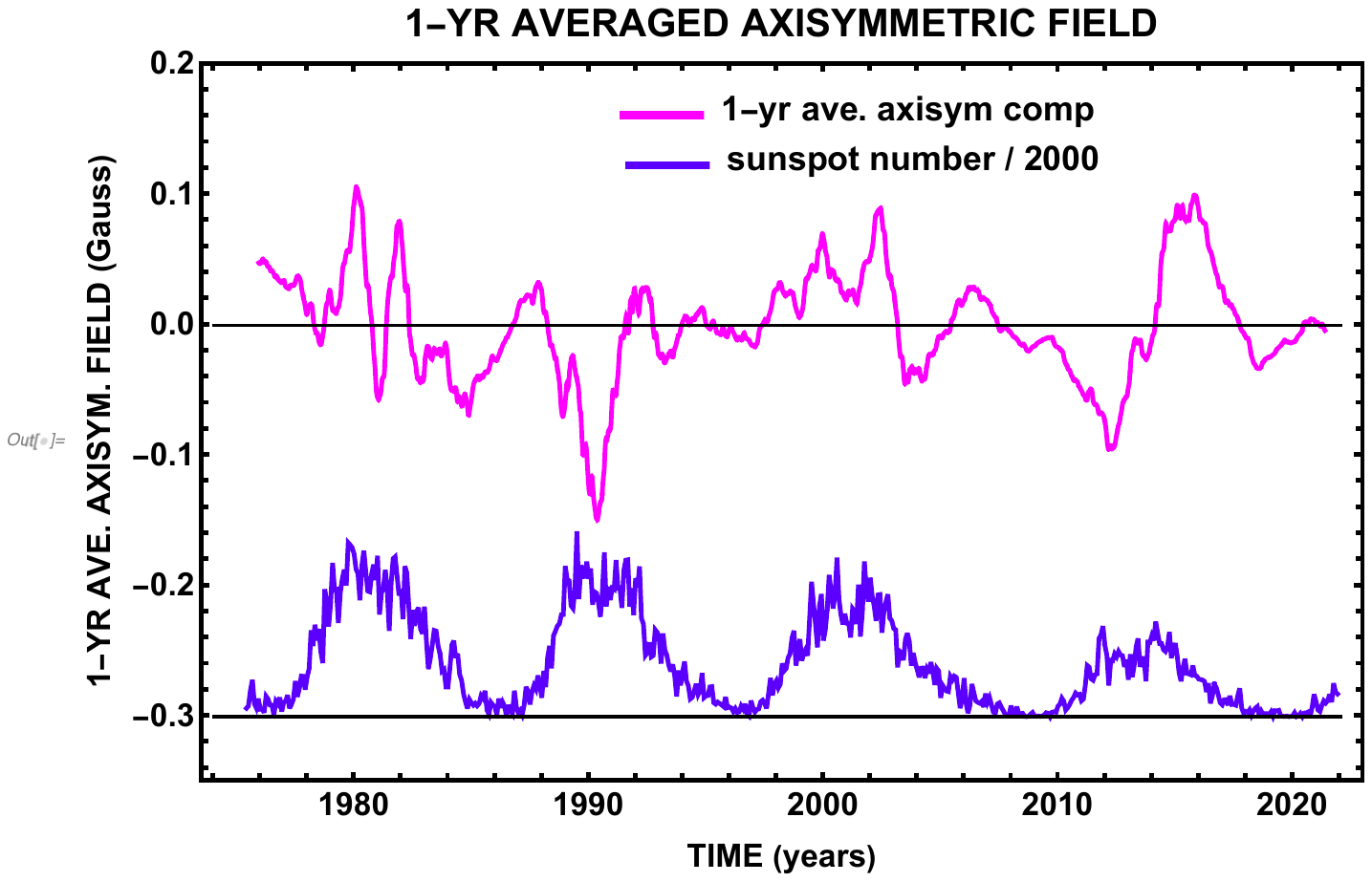}}}
\caption{(top) 27-day running average of the WSO mean field (red), compared with the
monthly averaged sunspot number from the Royal Observatory of Belgium (SILSO) (blue).
(bottom) 365-day moving average of the red curve, showing the residual $Y_{2}^{0}$
component of the axisymmetric field (purple) after the annually varying $Y_{1}^{0}$
and $Y_{3}^{0}$ contributions have been removed.
\label{fig:fig13}}
\end{figure}
\noindent
by 2000 and shifted downward by 0.55 units on the vertical scale.
At first glance, the red curve seems to be a meandering collection of relatively uninteresting
noisy wiggles.  However, closer inspection reveals a rough trend for the signal to be
negative during the early years of sunspot maximum in sunspot cycles 22 and 24 and
positive during the declining phase of those cycles.  The polarities are reversed during the
odd-numbered cycles 21 and 23.  In addition, there are well defined annual variations during
the 1976, 1986, and 2019 sunspot minima, which are presumably the most visible indications
of the $B_{0}$-dependence of the axisymmetric component of the mean field as Earth orbits
the Sun during the year. 

These $B_{0}$-induced variations are removed in the bottom panel of Figure~13,
whose purple curve is the 365-day running mean of the $<B_{m}>$ values associated with
the red curve in the upper panel.  We suppose that this purple curve indicates the $Y_{2}^{0}$
contribution given by ${\rho}_{20}I_{20}=-0.084{\rho}_{20}$ in Eq(17)\footnote{Note that
$Y^{0}_{2}$ has the sign of its polar region, which is negative when its equatorial region is positive, according to the definition $P_{2}(\cos{\theta})=(1/2)(3\cos^{2}{\theta}-1)$.  This accounts for the negative sign of $I_{20}$ in Table 1.}.  Thus, the mean field reduces the contribution of the
$Y_{2}^{0}$ component by a factor of about 12 and has a negative value corresponding
to the sign of the equatorial part of a positively directed quadrupole.

The temporal profile of the purple curve in Figure~13
is similar to the profile of the $Y_{2}^{0}$ component calculated from spatially
resolved observations at both WSO and MWO (but not shown here).  The main difference occurred
during 1992-1996 when the spatially resolved measurements give a much larger positive field
than the mean field in Figure~13.  This stronger field would have strengthened the rough,
alternating-polarity rule noted from the red curve in the top panel of Figure~13.

As pointed out previously by
\cite{2011ApJ...736..136W} and \cite{2012ApJ...755..135R}, this polarity rule reflects the tendency
of equatorial flux to originate from the leading parts of active regions together with the
greater activity in the southern hemisphere than in the northern hemisphere 
during the declining phases of sunspot cycles 21-24.  Likewise, the
northern hemispheres were more active during the rising phases of those cycles.   However,  \cite{2011ApJ...736..136W} also found that this alternating-polarity rule broke down during the very
high activity of sunspot cycle 19 when the northern hemisphere tended to be more active than the southern hemisphere throughout the cycle.  (See also Figure~1 of \cite{1977ApJS...33..391W} who
found no systematic variations during 1874-1971\footnote{The reader may have to go to
the hardcopy edition of this paper because the figure was a large foldout that was not
scanned into the online edition.}.)

The annually varying part of the axisymmetric field is shown by the purple curve in Figure~14.
\begin{figure}[h!]
\centerline{
  \fbox{\includegraphics[bb=110 420 500 671,clip,width=0.95\textwidth]
 {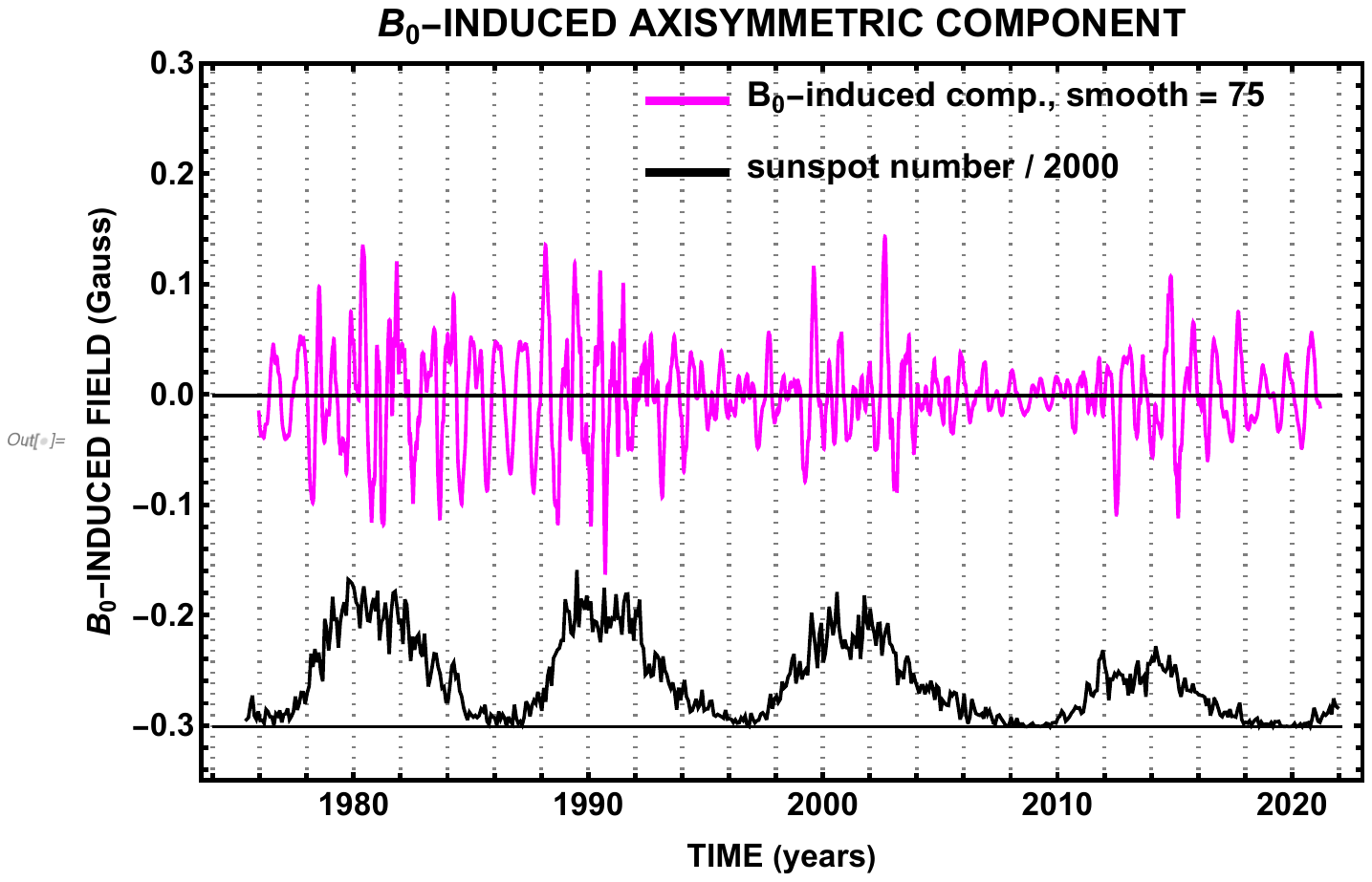}}}
\caption{The axisymmetric field, $<B_{m}>$, with its $Y_{2}^{0}$ component
removed and a 75-day smoothing applied to the remainder.  The resulting purple curve
shows annular modulation, especially during the 1976, 1986, and 2019 sunspot minima.
The black curve shows the monthly averaged sunspot number from the Royal
Observatory of Belgium (SILSO). 
\label{fig:fig14}}
\end{figure}
\noindent
We obtained this curve by subtracting the 1-year averaged field (purple curve in the bottom
panel of Figure~13) from the total axisymmetric field (red curve in the top panel of Figure~13),
and then taking a 75-day running average to remove the noise.
Referring to Eq(17), we expect this $B_{0}$-induced field to be
$B_{0}({\rho}_{10}J_{10}+{\rho}_{30}J_{30})$, where $B_{0}$ is given by
\begin{equation}
B_{0}(t)~{\approx}~0.126\sin \left \{ 2{\pi} \left ( \frac{t-157}{365} \right ) \right \}
\end{equation}
whose amplitude vanishes on day-of-year 157 (June 6) and reaches
0.126 rad (7.25$^{\circ}$) on day-of-year 249 (September 6),
and where
$J_{10}=+0.244$ and $J_{30}=-0.093$, as given in Table~2.  The vertical meandering
of the $Y_{2}^{0}$ component,
that was present in the top panel of Figure~13, is clearly gone.  The annual
variation is strongly visible around the 1976, 1986, and 2019 sunspot minima, but  only
weakly visible around the 1997 and 2009 minima.  Stronger non-periodic bursts occur
during the intervening sunspot-maximum intervals.

Faint dotted lines have been drawn at the even-numbered years and the panels have
been enlarged to help show the phase of the annual variation.  For example,
during 1976 and 1977, the positive peaks occurred in the fall and the negative peaks
occurred in the spring, as expected for a positive axisymmetric field.  In contrast, during
1985-1987, the sharp negative peaks occurred in the fall and the blunted positive peaks
occurred in the spring, consistent with a negative axisymmetric field.  This means
that the $B_{0}$-induced axisymmetric component of the mean field changed its sign
from plus to minus in going from sunspot cycle 21 to 22, in agreement with the signs of the axisymmetric dipole and the Sun's polar magnetic field.  The amplitude of the $B_{0}$-induced
component was much weaker during the 1997 and 2009 solar minima, but the
alternation of signs was still detectable.  Then in 2019, the field was strong again, and
its sign was positive as expected for the continued 11-yr alternation of polarity.
Thus, during the past 5 sunspot minima from 1976 to 2019, the $B_{0}$-induced
component of the mean field reversed its polarity in phase with the polarity of the
Sun's polar magnetic field and the Sun's axisymmetric dipole component.

Figures~15 and 16 provide more graphic displays of this phase alignment.
In Figure~15, plots of the WSO polar
\begin{figure}[h!]
 \centerline{
 \fbox{\includegraphics[bb=110 420 500 671,clip,width=0.95\textwidth]
 {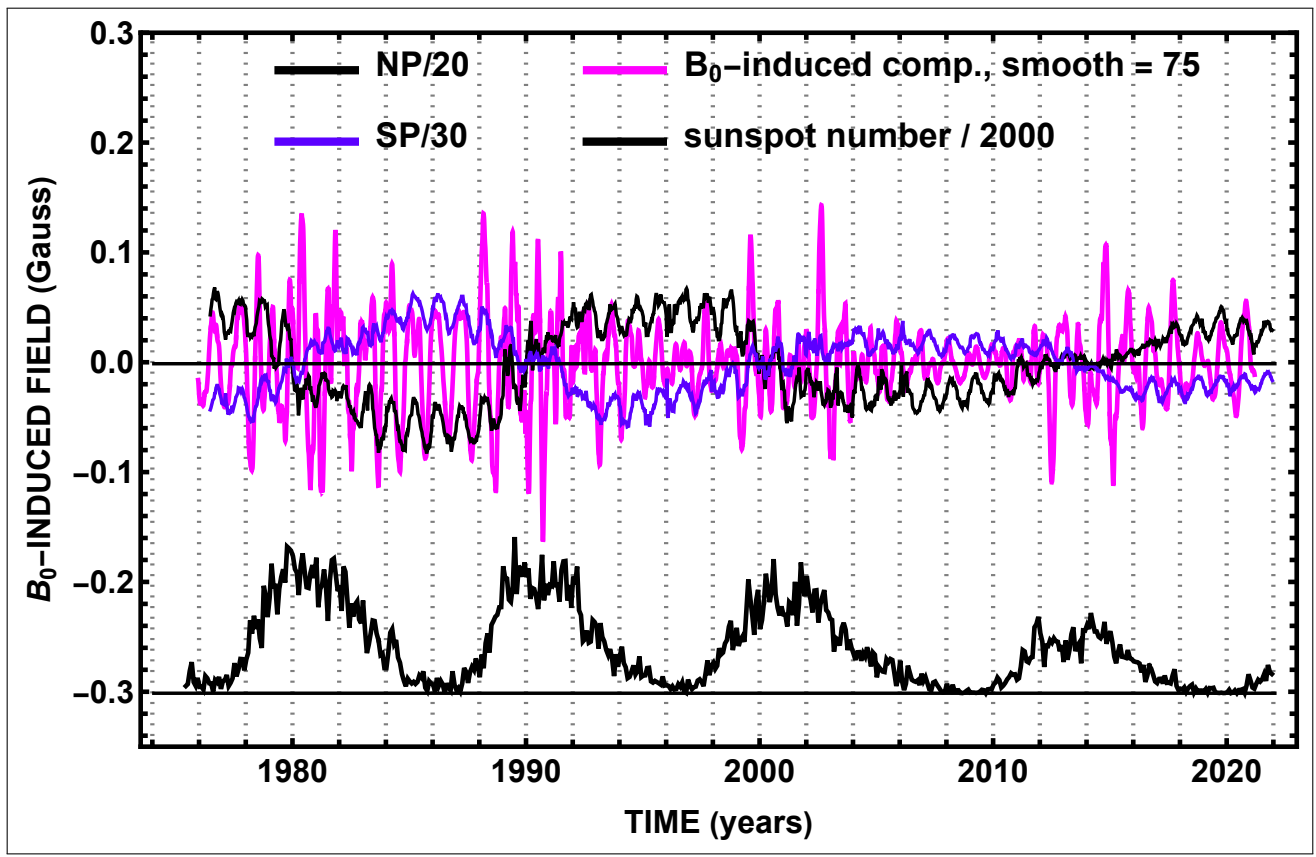}}}
\caption{WSO measurements of the Sun's north (black) and south (blue) polar
fields, superimposed on the smoothed plot of the $B_{0}$-induced part of the
axisymmetric field (purple), and compared with the monthly-averaged
sunspot number (black).  Annual variations of the purple curve are in phase with the
annual variations of the polar fields, but the polar fields had to be divided
by separate factors of 20 and 30 to improve the agreement between the long-term
envelopes of these curves. 
\label{fig:fig15}}
\end{figure}
\noindent
field strengths are superimposed on the $B_{0}$-induced part of the axisymmetric mean
field (shown in purple again).  To obtain the best overall agreement with the envelope
of the mean field, I reduced the north and south polar field strengths by factors of
20 and 30, respectively, before plotting them.  Now, the envelopes agree fairly well in the
years around sunspot minimum, but not
around sunspot maximum when the polar fields were reversing and the mean field had
several large bursts.  A detailed inspection of these
overlapping curves shows that the annual variations of the polar fields are in sync with
the annual variations of the $B_{0}$-induced field.  The north polar field
and its mean-field ripple reached their greatest absolute magnitudes in the fall of each year,
and the south polar field and its corresponding mean-field ripple reached their greatest
magnitudes in the spring.
 
Figure~16 provides another comparison between the $B_{0}$-induced mean field
and the polar fields.  In this case, 
\begin{figure}[h!]
 \centerline{
 \fbox{\includegraphics[bb=110 420 500 671,clip,width=0.95\textwidth]
 {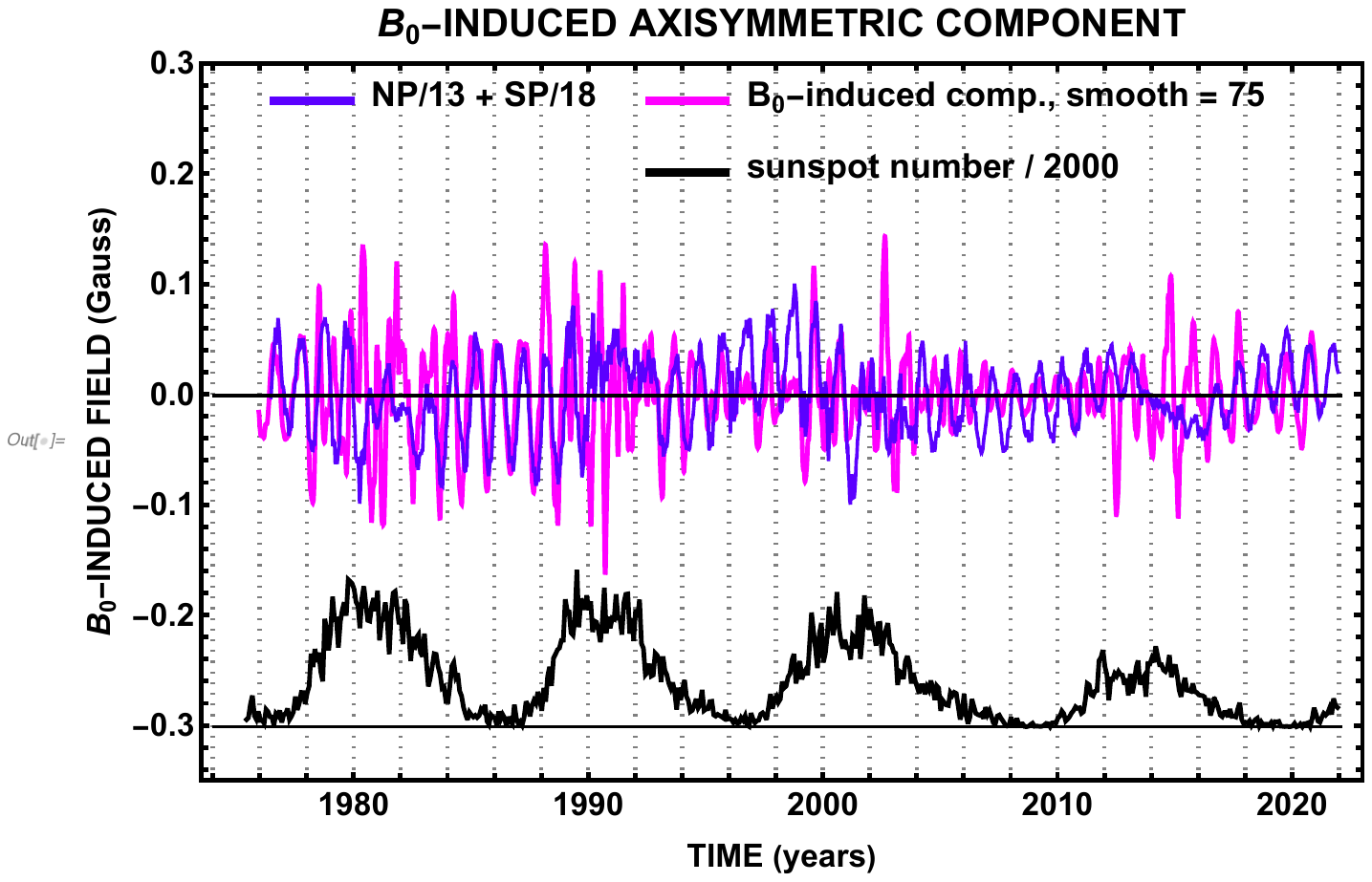}}}
\caption{Weighted sum of WSO north and south polar fields (blue), superimposed on
the $B_{0}$-induced part of the axisymmetric field (purple), and compared with the monthly-averaged sunspot number (black) during cycles 21 - 24.  Annual variations of this sum approximately match
those of the purple curve, except in the years around sunspot maximum when the polar fields reverse and the purple curve has large spikes.
\label{fig:fig16}}
\end{figure}
\noindent
the weighted values of the north and south polar fields are
added, to form a single oscillating blue curve  similar to the overlapping purple plot of
the mean field variations.  Also, the weighting is changed
slightly so that the contribution of the north pole is increased by a factor of
$18/13~{\approx}~1.4$, instead of the factor of 30/20 = 1.50 that was used in Figure~15. 
The overall trend is essentially the same as we found in Figure~15 with consistent
agreement in phase and fair agreement in magnitude, except in the years around sunspot
maximum when the polar fields changed sign and the mean field had several large purple spikes.

These graphical comparisons between the $B_{0}$-induced component
of the mean field and the amplitude of the polar field clearly indicate
that these two fields are in phase and that they reverse their polarities together
from one sunspot cycle to the next.  We might expect this phase synchronization
because both of these axisymmetric fields depend strongly on $Y_{1}^{0}$, which reverses
the polarity of its contribution from one sunspot cycle to the next.

However, the $B_{0}$-induced field and the polar field have differences that might cause
their amplitudes to differ.   The $Y_{3}^{0}$ component contributes to both fields, but with
opposite signs.  We have seen in Eq(17), that $Y_{3}^{0}$  makes a negative contribution
to the $B_{0}$-induced mean field because the coefficient, $J_{30}$, is negative.  On the other
hand, the topknot character of the polar field requires a weighted sum
of the $Y_{1}^{0}$ and $Y_{3}^{0}$ contributions (and smaller contributions from
$Y_{5}^{0}$ and $Y_{7}^{0}$) in order to strengthen the field at the poles and weaken
the field around the equator
\citep{1978SoPh...58..225S,1984SoPh...92....1D,1989SoPh..124....1S,1989SoPh..119..323S}.
Also, the north and south polar fields often differ in strength due to symmetry-breaking
contributions from harmonic components of even $l$, especially the $Y_{2}^{0}$
component.  As shown in the second column of Table~2, these even-order components do
not contribute to the $B_{0}$-induced axisymmetric component of the mean field.  This
explains why we had to equalize the strengths of the polar fields in order to
bring their plots into better agreement with the envelope of the $B_{0}$-induced
axisymmetric field in Figures~15 and 16.

\section{Summary and Discussion}
In this paper, I have regarded the Sun as an unresolved source of light like that from a distant
star, and calculated the transmission that this `mean-field filter' would have for the spherical harmonic components, $Y_{l}^{m}$, of the field.  This transmission depends on the mode
numbers, $l$ and $m$, and $B_{0}$, the observer's latitude in the star's polar coordinate
system.  The transmissions fell into three separate classes, proportional    
to $\cos^{2}B_{0}$, $(\sin2B_{0})/2$, and $\sin^{2}B_{0}$, as given in Tables 1-3, respectively,
and described in Eqs(5) and (7).  For the Sun, $B_{0}$ varies between $-7.^{\circ}25$ and
$+7.^{\circ}25$ during the year, so that $\sin^{2}B_{0}$ is small and Table 3 can be neglected.
In that case, we expect the axisymmetric part of the mean field to originate from the
$Y_{2}^{0}$ component and the first few harmonic components of odd $l$
(mainly $Y_{1}^{0}$
and $Y_{3}^{0}$), which are related to the polar fields and are coupled to the mean field
\textit{via} $B_{0}$.  On the other hand, we expect the  non-axisymmetric part of the mean field
to originate from the $Y_{1}^{1}$ and $Y_{2}^{2}$ components with occasional small contributions from the $Y_{3}^{3}$ and $Y_{3}^{1}$ components.

For our observations of the Sun, $B_{0}$ is small and the contribution of the $Y_{2}^{0}$
component comes from its equatorial band, which is negative for a positively defined
$Y_{2}^{0}$ component.  This is why the sign of $I_{20}$ is negative in Table~1.
On the other hand, for a distant star observed nearly head-on, $B_{0}$ would be
${\sim}90^{\circ}$, and the $Y_{2}^{0}$ contribution would come from one of its two polar regions of positive polarity.  This is why the sign of $K_{20}$ is positive in Table~3.

I applied these ideas to the 46-year series of WSO daily measurements of the Sun's
line-of-sight mean field, first plotting the 27-day running average of the rms deviation
of the mean field from its average value.  This rms variation provided a measure of the power
in the non-axisymmetric
components of the field, and showed peaks already familiar to us from plots of the occurrence
of coronal inflows, of the Sun's open flux and interplanetary magnetic field, of the Sun's equatorial dipole and quadrupole field, of the non-axisymmetric component of the Sun's source-surface
magnetic field, and of the mean field represented by a running 27-day average of its
maximum-minus-minimum values \citep{2014ApJ...797...10S, 2015ApJ...809..113S}.
In retrospect, this similarity of plots should be no surprise because they all show
peaks corresponding to the envelope of the non-axisymmetric component of the
Sun's large-scale field as it is modulated by solar rotation, analogous to the audio
component of a high-frequency radio wave.

The second step was to take the Fourier transform of the mean-field measurements and
display power as a function of frequency.  The result was a series of peaks at frequencies corresponding to the 27-day synodic solar rotation period and its first two harmonics.  In
addition, each peak showed fine structure at slightly longer periods, corresponding to the
Sun's rotation at higher latitudes.  By inverting this Fourier transform using broad windows
around these three peaks, I obtained a relatively noise-free version of the non-axisymmetric
mean field.  In addition, temporal plots of the $m=1$, $m=2$, and $m=3$ components of
the non-axisymmetric field were obtained by selecting each of the three windows separately.
By comparing these plots with the temporal plot of their sum, we learned that nearly all of the
large peaks were provided by the combination of $m=1$ and $m=2$ plus a few residual contributions of $m=3$.  In other words, the mean field is dominated by the $Y_{1}^{1}$
and $Y_{2}^{2}$ components of the field, consistent with our analysis of Eq(16).

The third step was to find the source of the fine structure in the power spectrum, and
in particular of the peak whose frequency corresponded to a period of ${\sim}$30 days.  This
narrow peak was particularly interesting because it had no
counterpart in the interplanetary sector structure inferred from Earth-based magnetometer measurements.  Those in-ecliptic measurements often show 28.5-day recurrence
patterns, but never 30-day patterns.

By inverting the Fourier transform of the mean field
through a narrow window around the 30-day peak, I found that most of the power originated
in 1989-1990 when photospheric magnetograms showed elongated patterns of magnetic fields migrating to high latitudes in the northern hemisphere.  This reminds us that
the mean-field is sampling large-scale magnetic patterns, which gradually rotate rigidly as supergranular diffusion and meridional flow carry their flux across latitudes
\citep{1987ApJ...319..481S,1987SoPh..112...17D,1998ASPC..154..131W}.
Thus, mean-field measurements give the pattern rate, which depends on the meridional
transport parameters as well as the rate of differential rotation, and not just the rotation rate
itself.  We are still left with the interesting quantitative question of why the 1989-1990 fields gave a rotation period of ${\sim}$30 days, rather than a longer period associated with
the high-latitude tail of the migrating stream or the shorter, ${\sim}$28.5-day period found
for so many other migrating streams.

If the answer lies in the strength of the active region that emerged at 35$^{\circ}$N latitude
in 1989, then we might ask if similar (or even stronger) active regions may have emerged at high latitudes in previous sunspot cycles, like cycles 18 and 19, that were more active than cycles 21 - 24.  A large northern-hemisphere pattern was visible in Ca II K-line maps and
Fe I 5250 \AA~ magnetograms obtained at the Mount Wilson Observatory during
Carrington rotations 1417 (8 August - 4 September, 1959) and 1419 (2-29 October 1959)
\citep{2011ApJ...730...51S}.  Another occurred in the southern hemisphere during
rotations 1259 (21 October -  17 November 1947) and 1261
(14 December 1947 - 11 January 1948).  Perhaps those migrating fields would have
produced $m=1$ sidebands with rotation periods of at least 30 days in the power
spectrum of the mean field.

The fourth step was to look at the axisymmetric component of the mean field.  This
component consisted of the $Y_{2}^{0}$ component plus the strongest $B_{0}$-induced
components of odd $l$, mainly $Y_{1}^{0}$ and $Y_{3}^{0}$.  Because $B_{0}$ varies
annually due to Earth's orbital motion around the Sun, it was possible to remove the
$B_{0}$-induced term, and display the annually averaged $Y_{2}^{0}$ component
separately.  Although not shown here, its temporal profile was similar to those obtained
from spatially resolved observations at both WSO and MWO, except for
the interval 1992-1996 when the spatially resolved measurements gave a much larger
positive field.  Resolving this discrepancy will be a challenge for the future.

Once the $Y_{2}^{0}$ component was found, it was then possible to extract the
$B_{0}$-induced part of the axisymmetric field.  Its annual variation was
in phase with the annual variation of the polar magnetic fields.  This means that the
$B_{0}$-induced field was oriented in the same direction as the polar field, reversing
its direction from one sunspot cycle to the next.   A remaining
puzzle is why the annual variations were strong around the 2019 sunspot minimum 
when the polar fields were weak, and why the annual variations were weak around
the 1997 minimum when the polar fields were stronger.

Finally, we note that this approach can be used to find spherical harmonic
components of the fields in other stars.  For example, the $B_{0}$-dependence
of the strengths of these harmonic components may complement asteroseismology determinations of the orientations of the rotational axes of these stars \citep{2003ApJ...589.1009G}.  Also, the appearance of low-frequency sidebands in the WSO spectra suggests that similar sidebands may occur in observations of other stars, providing information about differential rotation and meridional flow in those stars.   This information may help to remove ambiguities in the inferences of
large scale convection and magnetic cycles in asteroseismology studies of Sun-like stars.

\begin{acknowledgments}
I am grateful to Phil Scherrer (WSO/Stanford) and Todd Hoeksema (WSO/Stanford) for
helpful comments about the WSO telescope and its data, and Yi-Ming Wang (NRL)
for numerous discussions of the Sun's large-scale magnetic field.  I am grateful to
Jack Harvey (NSO/LPL/UA) for helping me find archived Carrington maps of observations
obtained with the Kitt Peak Vacuum Telescope.  This work evolved from a talk
that I gave at an LPL/UA Heliophysics Group Zoom session where Joe Giacalone,
Jack Harvey, and John Leibacher provided useful comments and ideas.  Wilcox Solar
Observatory data used in this study were obtained \textit{via} the web site http://wso.stanford.edu
courtesy of J.T. Hoeksema. NSO data were acquired by SOLIS instruments operated by
NISP/NSO/AURA/NSF.  Sunspot numbers were obtained from WDC-SILSO, Royal Observatory of Belgium, Brussels.  I am grateful to the referee for several helpful comments, including
one about limb darkening which motivated the calculations in the Appendix.

\end{acknowledgments}





\appendix
\section{Limb Darkening}
In Section 2, we converted the integral of the line-of-sight field over the
(flat) solar disk to a surface integral of the radial field over the visible hemisphere.
As shown in Eq(3), this conversion introduced two factors of ${\mu}$ into the integrand,
causing the surface integral of $B_{r}$ to be weighted toward the disk center by a factor
of ${\mu}^2$.  However, we did not include the natural weighting that is produced by the limb darkening of the Sun and stars in the visual region of the spectrum.  This limb darkening occurs because light from disk center originates in deeper, hotter, and brighter layers than light from positions toward the limb.  As described by \cite{1977SoPh...52D...6S}, the central weighting of the WSO measurements is mainly due to this natural limb darkening plus a contribution of diffraction
from the entrance slit of the spectrograph.

It is relatively easy to include limb darkening in our equations for the mean line-of-sight
magnetic field.  We simply insert the limb darkening intensity profile, $I({\mu})/I(0)$, next to the
${\mu}^2$ in the integrand of Eq (3).  With the definition $F({\mu})=I({\mu})/I(0)$, Eq(3)
becomes 
\begin{equation}
B_{m}~=~\int{B_{r}}{\mu}^{2}F({\mu})dA_{surf}/{\pi}R^2.
\end{equation}
Then, we replace ${\mu}$ by $\sin{\theta}\cos{\phi}\cos{B_{0}}+\cos{\theta}\sin{B_{0}}$,
and perform the integration over the variables ${\theta}$ and ${\phi}$ as indicated in
Eqs(4) and (5).  If $F({\mu})$ is a simple first- or second-order polynomial in ${\mu}$, one can
do the integration analytically, but if the profile is more complicated, then it might be more
convenient to integrate numerically.

Let's begin by considering some of the limb darkening relations that are used to describe
the Sun and stars.  Classically, the natural limb darkening has been described by a linear
function of ${\mu}$ of the form
\begin{equation}
 \frac{I({\mu})}{I(0)}~=~1-{\gamma}(1-{\mu})=(1-{\gamma})+{\gamma}{\mu}.
\end{equation}
Here, ${\gamma}$ is a wavelength-dependent limb-darkening coefficient that indicates how
dark the limb is relative to the disk center at that wavelength.  For the so-called gray atmosphere
in the Eddington approximation \citep{1990soas.book.....F,1906WisGo.195...41S},
${\gamma}=0.6$ and $I({\mu})/I(0)=0.4+0.6{\mu}$.
According to \cite{2018A&A...616A..39M}, \citeauthor{1950HarCi.454....1K}'s \citeyearpar{1950HarCi.454....1K} quadratic limb darkening relation of the form
 \begin{equation}
 \frac{I({\mu})}{I(0)}~=~1-c_{1}(1-{\mu})-c_{2}(1-{\mu})^2
 \end{equation}
is the most commonly used profile in modern exoplanet studies.  Stellar limb darkening is important in these studies because it affects the light curve produced by the transiting exoplanet and therefore the accuracy with which the exoplanetary radius can be determined.  Consequently, other profiles
are sometimes used to match the high-precision light curves of transiting
exoplanet systems \citep{2007ApJ...655..564K,2015MNRAS.450.1879E}.  In particular,  \cite{2018A&A...616A..39M}  considered another two-parameter limb-darkening relation
of the form
\begin{equation}
\frac{I({\mu})}{I(0)}~=~1-c(1-{\mu}^{\alpha}).
\end{equation}
 
 Although any of these limb darkening relations can be substituted into Eq(A1) above, I will
illustrate the procedure using the linear relation given by Eq(A2).  In this case, $B_{m}$ can be obtained from proportionate parts of Eq(4) and of the modified version of Eq(4) when an
extra factor of ${\mu}$ is included in its integrand.  This latter quantity, which I call $B_{\mu}$
to distinguish it from $B_{m}$, is given by
 \begin{equation}
 B_{\mu}~=~(1/{\pi})\int_{-{\pi}/2}^{{\pi}/2}\int_{0}^{\pi}{B_{r}}(\sin{\theta}\cos{\phi}\cos{B_{0}}+\cos{\theta}\sin{B_{0}})^{3}\sin{\theta}d{\theta}d{\phi}.
 \end{equation}
The third-power expansion of this binomial expression for ${\mu}$ gives
 four terms with factors of $\cos^{3}{B_{0}}$, $3\cos^{2}{B_{0}}\sin{B_{0}}$, $3\cos{B_{0}}\sin^{2}{B_{0}}$, and $\sin^{3}{B_{0}}$, respectively, instead of the three terms that we obtained
from the second-power expansion of that binomial in Eq(4).

It would be easy to evaluate all four integrals and put the results in tables as we did
in the main text.  However, to estimate the effect of limb darkening on the mean field of the Sun,
it is sufficient to expand the $B_{0}$-dependent factors in powers of $B_{0}$ and retain only
the zeroth-order and first-order terms.  This means that we need to keep only the $\cos^{3}{B_{0}}$ and $3\cos^{2}{B_{0}}\sin{B_{0}}$ factors, which reduce to 1 and $3B_{0}$, respectively.  Then, the limb darkened mean field becomes
\begin{equation}
B_{m}~=~(1-{\gamma})(I_{lm}+B_{0}J_{lm})~+~{\gamma}(I_{lm}^{\mu}+B_{0}J_{lm}^{\mu}),
\end{equation}
where the terms that originate from Eq(A5) are given by
\begin{subequations}
\begin{align}
I_{lm}^{\mu}~=~\frac{N_{lm}}{{\pi}} \int_{-1}^{1}P_{l}^{m}(x)(1-x^2)^{3/2}dx
\left [ \frac{12 \cos{(m{\pi}/2)}}{(m+1)(m-1)(m+3)(m-3)} \right ],\\
J_{lm}^{\mu}~=~3\frac{N_{lm}}{{\pi}}  \int_{-1}^{1}P_{l}^{m}(x)(1-x^2)xdx
\left [ \frac{-4 \sin{(m{\pi}/2)}}{m(m+2)(m-2)} \right ].
\end{align}
\end{subequations}
Next, we rearrange Eq(A6) as a power series in $B_{0}$ to obtain
\begin{equation}
B_{m}~=~\{(1-{\gamma})I_{lm}+{\gamma}I_{lm}^{\mu}\}~+~B_{0}
\{(1-{\gamma})J_{lm}+{\gamma}J_{lm}^{\mu}\}.
\end{equation} 
I used the conventional Mathematica software to evaluate $I_{lm}^{\mu}$ and
$J_{lm}^{\mu}$ for $l$ and $m$ in the range $0-7$, again reversing the signs of the
odd-m entries to be consistent with the \cite{JE_45}-convention.  Then I set
${\gamma}=0.6$, and combined these results with the values of $I_{lm}$ and $J_{lm}$ given in Tables 1 and 2 to obtain the limb darkened intensities for the gray atmosphere in the Eddington approximation.  The results are given in Tables 4 and 5.
\begin{table}[h!]
\caption{Limb-Darkened Elements of $B_{lm}$ \{zeroth-order in $B_{0}$\}}
\begin{center}
\begin{tabular}{c c c c c c c c c c}
\hline\hline
$l/m$ & 0&1 & 2 & 3 & 4 & 5 & 6 & 7 \\[0.5ex]
\hline\
0 &+0.160 & \\[1.5ex]

1 & 0 & +0.152  \\[1.5ex]

2 & -0.081 & 0 & +0.099	\\[1.5ex]

3 & 0 & -0.033 & 0 & +0.043	 \\[1.5ex]

4 & +0.006 & 0 & -0.006 & 0 & ~~+0.008~~  \\[1.5ex]

5 & 0 & -0.001 & 0 & +0.002 &0 & -0.002  \\[1.5ex]

6 & +0.001 & 0 & -0.001 & 0 & +0.001 & 0 & ~~-0.001~~  \\[1.5ex]

7 & 0 & -0.000 & 0 & +0.000 & 0 & -0.000 & 0 & +0.001  \\[1.5ex]
\hline 
\end{tabular}
\end{center}
\end{table} 
\begin{table}[h!]
\caption{Limb-Darkened Elements of $B_{lm}$ \{first-order in $B_{0}$\}}
\begin{center}
\begin{tabular}{c c c c c c c c c}
\hline\hline
$l/m$ & 0&1 & 2 & 3 & 4 & 5 & 6 & 7 \\[0.5ex]
\hline\
0 & 0 & \\[1.5ex]

1 & +0.215 & 0  \\[1.5ex]

2 & 0 & +0.198 & 0	\\[1.5ex]

3 & -0.114 & 0 & +0.104 & ~~0~~	 \\[1.5ex]

4 & 0 & -0.026 & 0 & +0.023 & 0  \\[1.5ex]

5 & -0.007 & 0 & +0.007 & 0 & -0.006 & ~~0~~  \\[1.5ex]

6 & 0 & -0.004 & 0 & +0.004 & 0 & -0.003 & 0  \\[1.5ex]

7 & -0.003 & 0 & +0.003 & 0 & -0.003 & 0 & +0.002 & ~~0~~  \\[1.5ex]
\hline 
\end{tabular}
\end{center}
\end{table} 

Comparing Table 1 and Table 4, we see that the elements have the same sign and nearly the
same magnitude for $l<4$.  For larger values of $l$, there are some sign differences,
especially for $l=4$ and $l=6$, but the magnitudes of these higher-order harmonic
components are less than 0.01 and can be neglected.  A comparison between Tables 2
and 5 gives a similar result.  This means that the same harmonic components contribute to
the mean field with or without limb darkening, but that the strengths of these contributions
differ slightly, depending on the values of ${\gamma}$ and $l$ (as we shall see next).

We can gain further insight by computing the relative differences between these limb-darkened
and non-limb-darkened intensities.  Doing this separately for the terms that are zero-order
and first-order in $B_{0}$, we obtain
\begin{equation}
(\frac{{\Delta}B}{B})_{0}~=~\frac{\{(1-{\gamma})I_{lm}+{\gamma}I_{lm}^{\mu}\}-I_{lm}}{I_{lm}}~=~-{\gamma}(1-\frac{I_{lm}^{\mu}}{I_{lm}})
\end{equation}
and
\begin{equation}
(\frac{{\Delta}B}{B})_{1}~=~\frac{\{(1-{\gamma})J_{lm}+{\gamma}J_{lm}^{\mu}\}-J_{lm}}{J_{lm}}~=~-{\gamma}(1-\frac{J_{lm}^{\mu}}{J_{lm}}).
\end{equation}
Thus, the zeroth-order change, $({\Delta}B/B)_{0}$, depends on the ratio $I_{lm}^{\mu}/I_{lm}$,
and the first-order change,  $({\Delta}B/B)_{1}$, depends on the ratio $J_{lm}^{\mu}/J_{lm}$.  And both changes are proportional to the limb darkening coefficient, ${\gamma}$.  Moreover, if we use Eqs(A7ab) and Eqs(11ab) to evaluate the ratios, $I_{lm}^{\mu}/I_{lm}$ and
$J_{lm}^{\mu}/J_{lm}$, we obtain the remarkable result that the values of these ratios
are the rational numbers 4/5, 15/16, and 48/35, for $l=1$, 2, and 3, respectively, independent of the values of $m$.  (Of course, this applies only for the non-zero values of $I_{lm}$ and $J_{lm}$
in Tables 1 and 2, respectively.) 

In other words, for the $Y_{1}^{0}$ and $Y_{1}^{1}$ harmonics, the ratios, $J_{10}^{\mu}/J_{10}$ and  $I_{11}^{\mu}/I_{11}$, are both equal to 4/5.  Likewise, for the $Y_{2}^{0}$, $Y_{2}^{1}$, and $Y_{2}^{2}$ harmonics, the ratios are all equal to 15/16.  And for the four harmonics with $l=3$,
the ratios are 48/35.  Subtracting these ratios from 1, we obtain 1/5, 1/16, and -13/35, as the
fractional differences in Eqs(A9) and (A10) for $l=1$, 2, and 3, respectively, before the limb darkening factor, ${\gamma}$, is applied.
 
 If we let ${\gamma}=0.6$, we obtain changes of -12\%, -3.75\%, and +22.2\%, for $l=1$, 2, and
 3, respectively, independent of the value of $m$.  Thus, for an ideal gray atmosphere in the
 Eddington approximation, the amplitudes of the $Y_{1}^{0}$ and $Y_{1}^{1}$ components
 decrease by 12\%.  The three components with $l=2$ decrease by only 3.75\%
 and the four relatively weak components with $l=3$ go in the opposite direction,
 increasing by 22.2\%.
 
 The nice aspect of Eqs(A9) and (A10)
 is that we can increase ${\gamma}$ to obtain the fractional changes for a greater amount
 of limb darkening in the violet part of the spectrum, or decrease ${\gamma}$ to obtain the smaller changes expected in the infrared.  We simply return to the fractions 1/5, 1/16, and -13/35 and
 multiply them by $-{\gamma}$.

I did this calculation to learn how solar limb darkening in the simple Eddington approximation
might affect the harmonic components of the mean field.  However, for more complex
limb darkening profiles and the more inclined rotational axes, that might occur in exoplanet or asteroseismic studies, one could relax the small-$B_{0}$ approximation used for the Sun,
and do the integration numerically for specific values of $B_{0}$.  We might expect limb
darkening to have a more complicated effect for a star whose rotational axis makes an
oblique angle to the line of sight.




\bibliography{mfield}{}
\bibliographystyle{aasjournal}



\end{document}